\documentclass[a4paper,11pt]{article}
\usepackage{jcappub}
\usepackage{aas_macros}
\usepackage{graphicx}
\usepackage{amsmath}
\usepackage{amssymb}
\usepackage{hyperref}
\usepackage{physics}
\usepackage[capitalize]{cleveref}

\usepackage{mathrsfs}

\makeatletter
\gdef\@fpheader{\mbox{}}
\makeatother

\newlength{\fullw}
\newlength{\halfw}
\newlength{\threefigw}
\newlength{\twofigw}
\newlength{\onefigw}
\newlength{\bigfigw}
\DeclareMathOperator{\erfc}{erfc}

\newcommand{\dirac}[1]{\delta\negthinspace\left(#1\right)}

\newcommand{\heaviside}[1]{\mathrm{\Theta}\!\left( #1 \right)}

\newcommand{\boldmathsymbol}[1]{{\ensuremath{\boldsymbol{#1}}}}

\newcommand{\sss}[1]{{\scriptscriptstyle{#1}}}
\newcommand{\efolds}{$e$-folds}
\newcommand{\efold}{$e$-fold}

\newcommand{\bx}{\boldmathsymbol{x}}

\newcommand{\uPl}{\mathrm{Pl}}

\newcommand{\uinf}{\mathrm{inf}}
\newcommand{\uqw}{\mathrm{qw}}

\newcommand{\uLT}{\mathrm{LT}}

\newcommand{\ufw}{\mathrm{fw}}

\newcommand{\usssPl}{\sss{\uPl}}

\newcommand{\usssLT}{\sss{\uLT}}
\newcommand{\umode}{\mathrm{mode}}
\newcommand{\zero}{{_0}}

\newcommand{\Mp}{M_\usssPl}

\newcommand{\Nqw}{N_\uqw}

\newcommand{\Nzero}{N_\zero}

\newcommand{\Nflat}{N_{\flat}}

\newcommand{\deltaN}{\var N}

\newcommand{\rvp}[1]{\bar{#1}}
\newcommand{\rv}[1]{\Delta#1}
\newcommand{\Pname}{P}
\newcommand{\Pof}[1]{\Pname\negthinspace\left(#1\right)}
\newcommand{\Pfw}[1]{\Pof{#1}}
\newcommand{\Prv}[1]{\rvp{\Pname}\negthinspace\left(#1\right)}
\newcommand{\Plt}[1]{\Pname_{\usssLT}\negthinspace\left(#1\right)}

\newcommand{\Frv}{\rvp{F}}

\newcommand{\Hinf}{H_\uinf}

\newcommand{\phiqw}{\phi_\uqw}

\newcommand{\phizero}{\phi_\zero}

\newcommand{\phiinf}{\phi_{\infty}}

\newcommand{\chizero}{\chi_\zero}
\newcommand{\chihat}{\hat{\chi}}
\newcommand{\chizerohat}{\chihat_\zero}

\newcommand{\Vzero}{V_\zero}

\newcommand{\calN}{\mathcal{N}}
\newcommand{\calI}{\mathcal{I}}

\newcommand{\eps}[1]{\epsilon_{#1}}

\newcommand{\zetahat}{\hat{\zeta}}
\newcommand{\zetafwhat}{\hat{\zeta}_{\ufw}}
\newcommand{\alphahat}{\hat{\alpha}}

\newcommand{\zetafw}{\zeta_{\ufw}}

\begin{document}

\title{Friction in Stochastic Inflation}

\author{Baptiste Blachier}
\author{and Christophe Ringeval}
\affiliation{Cosmology, Universe and Relativity at Louvain (CURL),
Institute of Mathematics and Physics, University of Louvain,
2 Chemin du Cyclotron, 1348 Louvain-la-Neuve, Belgium}

\emailAdd{baptiste.blachier@uclouvain.be}
\emailAdd{christophe.ringeval@uclouvain.be}

\date{\today}

\abstract{We solve time-reversed stochastic inflation in the
  semi-infinite flat potential with a constant drift term and derive
  an exact expression for the probability distribution of the
  curvature fluctuations. It exhibits exponential decaying tails which
  contrast to the Levy-like power law behaviour encountered without
  friction. Such a non-vanishing drift acts as a regulator for the
  conventional ``forward'' stochastic $\delta N$-formalism, which is
  otherwise ill-defined in the unbounded and flat potentials typical
  of plateau models of inflation. This setup therefore allows us to
  compare the curvature distribution derived from both approaches,
  reverse and forward in time. Up to similar exponential tails, we
  find quantitative differences. In particular, in the classical-like
  limit of very large drift, the tails become Gaussian but only in the
  time-reversed picture. As a toy model of eternal inflation, we
  finally discuss the case of negative drift in which inflation never
  ends for many field trajectories. The forward approach becomes
  pathological whereas the reverse formalism gives back a finite
  curvature distribution with always exponential tails. All these
  differences end up being related to the very definition of the
  background which is ambiguous when a classical trajectory does not
  exist.}

\keywords{Time-reversed Stochastic Inflation, Semi-infinite potential, Constant friction}

\maketitle

\section{Introduction}
\label{sec:intro}

Within the inflationary paradigm, cosmological perturbations are of
quantum mechanical origin, generated during the semi-classical regime
of Cosmic Inflation~\cite{Starobinsky:1979ty, Starobinsky:1980te,
  Sato:1980yn, Guth:1980zm, Linde:1981mu, Albrecht:1982wi,
  Linde:1983gd, Mukhanov:1981xt, Mukhanov:1982nu, Starobinsky:1982ee,
  Guth:1982ec, Hawking:1982cz, Bardeen:1983qw}. In this regime, the
space-time is classical and in a quasi-exponential accelerated
expansion: the inevitable outcome of a self-gravitating scalar field
having a non-vanishing potential energy~\cite{Chowdhury:2019otk,
  East:2015ggf, Clough:2017efm, Joana:2020rxm, Joana:2022uwc,
  Joana:2024ltg}. On top of this homogeneous background, the
field-metric system is experiencing small quantum fluctuations which
are the seeds of the large scale structures of the
universe~\cite{Mukhanov:1987pv, Mukhanov:1988jd, Stewart:1993bc}. As
of today, this mechanism is one of simplest and most favoured
explanation of the current cosmological observations and the preferred
potential shapes are
plateau-like~\cite{Planck:2018jri,EIopiparous,Martin:2024qnn}.

Quantum diffusion refers to another possible regime of Cosmic
Inflation, the one in which quantum fluctuations dominate over the
classical dynamics. This may occur either when quantum fluctuations
are large, i.e., when the Hubble parameter during inflation approaches
the Planck scale, or, when the classical dynamics becomes irrelevant,
as in nearly flat potentials. The expanding space-time can become
strongly inhomogeneous and the concept of a classical background
disappears~\cite{Vilenkin:1983xq, 1986PhLB..175..395L,
  Winitzki:2006rn, Winitzki:2008yb, Creminelli:2008es, Martin:2016iqo}. Cosmological
observations provide strong bounds on the inflationary dynamics, it
has to be semi-classical on the observed length scales. However, it is
still possible that quantum diffusion has something to do with the
very large or the very small length scales~\cite{Ringeval:2010hf,
  Kleban:2012ph, Glavan:2017jye, Ringeval:2019bob,
  Blachier:2023ooc}. Although not new, the topic has received renewed
interest in the recent years as a mechanism to produce primordial
black holes (PBHs)~\cite{Carr:2023tpt}. Indeed, a small flat domain
within the field potential $V(\phi)$, located towards, or after, the
end of the semi-classical regime is sufficient to create small scale
curvature fluctuations which are large enough to collapse into
PBHs~\cite{Vennin:2020kng, Tada:2021zzj, Tomberg:2022mkt, Raatikainen:2023bzk,
  Stamou:2023vft, Stamou:2023vwz, Animali:2024jiz,Tomberg:2025fku,
  Animali:2025pyf, auclair_2025_15235932, Raatikainen:2025gpd}.

When quantum fluctuations dominate, it is still possible to
(approximately) describe the dynamics of the field and of the
space-time by using the so-called stochastic inflation
formalism~\cite{Starobinsky:1986fx, Goncharov:1987ir, Nambu:1987ef,
  Kandrup:1988sc, Nakao:1988yi, Starobinsky:1994bd, Linde:1993xx,
  Salopek:1998qh}. It consists in considering the cumulated
contribution of all sub-Hubble quantum fluctuations as a Gaussian
white noise $\xi$ acting on the super-Hubble dynamics of a
coarse-grained inflaton field $\phi$. For not too steep potentials,
and out of the ultra slow-roll unstable regime, $\phi$ can be shown to
follow a Langevin equation~\cite{Grain:2017dqa}
\begin{equation}
\dv{\phi}{N} = -\dfrac{1}{3 H^2} \dv{V}{\phi} + \dfrac{H}{2\pi} \xi(N),
\label{eq:langevin}
\end{equation}
where we have taken Planck units ($\Mp=1$). Here $H(\phi)$ stands for
the Hubble parameter during inflation, $N=\ln a$ is the number of
{\efolds}, $a$ being a scale factor for the
Friedmann-Lema\^{\i}tre-Robertson-Walker (FLRW) metric. Up to the
noise term, this equation is manifestly the same as the one of a
homogeneous slow rolling scalar field but this is a non-trivial
property coming from the fact that the time variable has been chosen,
on purpose, to be the number of {\efolds}. Indeed, it is the only
choice consistent with quantum field theory
expectations~\cite{Finelli:2008zg, Vennin:2015hra}. Stochastic realisations of the
field trajectory drawn from \cref{eq:langevin} are now mapping the
quantum histories of the field-metric system.

Within the so-called separate universe picture, stochastic
realisations of \cref{eq:langevin} can also be associated with
different regions of space-time which, due to the coarse-graining, are
necessarily separated by super-Hubble distances. For super-Hubble
scalar inhomogeneities, one can always recast the metric as
$\dd{s}^2=-\dd{t}^2+a^2(t)\delta_{ij}e^{2 \zeta(\bx)}\dd{x}^i\dd{x}^j$
where $\zeta(\bx)$ is a conserved quantity~\cite{Lyth:2004gb,
  Rigopoulos:2004gr,Creminelli:2004yq, Langlois:2005ii}. As a result,
defining the {\efold} number $N(\bx,t) \equiv a(t)e^{\zeta(\bx)}$ in
each of these comoving regions yields to seemingly-like homogeneous
equations even though inhomogeneities are present. The
$\deltaN$-formalism is precisely based on this
observation~\cite{Sasaki:1995aw, Sasaki:1998ug, Wands:2000dp,
  Lyth:2004gb, Lyth:2005fi}. Starting from a coordinate choice
ensuring an initial flat slicing and evolving the field up to a
uniform energy density hypersurface, curvature fluctuations on
super-Hubble scales are given by $\zeta(\bx)= N(\bx,t) - \Nflat$,
where $\Nflat$ stands for a reference unperturbed number of
{\efolds}. The stochastic $\delta N$-formalism generalises these
results to stochastic inflation~\cite{Fujita:2013cna, Fujita:2014tja,
  Ando:2020fjm, Mizuguchi:2024kbl, Launay:2024qsm}. Because the field
trajectories are stochastic, the number of {\efolds} $\calN$ elapsed
from an initial field value $\phizero$ to $\phiqw$, the field value on
the ``quantum wall'' where quantum diffusion ends, is a stochastic
variable and so it is for the ``forward'' curvature perturbation
$\zetafw\equiv \calN - N_{\flat}$. Combined with \cref{eq:langevin},
the stochastic $\delta N$-formalism allows for the determination of
the statistical properties of $\zetafw$. The formalism has been
applied in the context of PBH formation, where it has been shown that
a short period of stochastic inflation occurring within a flat (or
quasi-flat) and bounded region of the potential, the so-called
``quantum well'', leads to an enhanced curvature probability
distribution having exponential tails instead of the typical Gaussian
statistics~\cite{Pattison:2017mbe, Ezquiaga:2019ftu, Figueroa:2020jkf,
  Figueroa:2021zah, Animali:2022otk, Calderon-Figueroa:2025dto}.

Let us remark that, in the aforementioned stochastic $\delta
N$-formalism, the number of {\efolds} is counted forward, from an
initial field state $\phizero$ to the end of quantum diffusion at
$\phiqw$. In spite of its name, it therefore shares more resemblances
with the so-called forward $\delta n$-formalism rather than with the
genuine classical $\delta N$-formalism~\cite{Cruces:2025typ}. Indeed,
curvature perturbations, as measured by observers attached to the end
of inflation hypersurfaces, are given by the fluctuations of the
number of {\efolds} counted backward, namely from the end of inflation
to the initial state~\cite{Sasaki:1998ug}. For instance, one can use
the (backward) formalism to recover that the squeezed bispectrum
measured by local observers exactly vanishes for single-field
inflation whereas the forward approach predicts that it scales as
$1-n_\mathrm{S}$~\cite{Tada:2016pmk}. In the cases where quantum
effects are sub-dominant, the differences remain small, but are they
when quantum diffusion is important? Another remark concerns the value
of $\Nflat$. In practical applications, stochastic effects usually
occur on top of a classical trajectory, and, in these situations,
$\Nflat$ is taken as the classical number of {\efolds}. An exception
to this case is the quantum well where the potential is exactly flat,
but bounded~\cite{Pattison:2017mbe}. In there, it seems natural to
take $\Nflat=\ev{\calN}$ which is a well defined quantity. Moreover,
it leads to $\ev{\zetafw}=0$, which is satisfactory as the space-time
over which quantum fluctuations are integrated out is considered to be
globally flat (see Refs.~\cite{Handley:2019anl,
  Letey:2022hdp,Vigneron:2024bfj} for non-vanishing curvature). Let us
notice that there are situations in which the choice of $N_{\flat}$ is
irrelevant, as for instance when one compares $\zeta$ between two
different locations. This is particularly true in the context of PBH
where the formation criteria are associated with the so-called
compaction function which involves differential operators acting on
$\zeta$~\cite{Shibata:1999zs, Musco:2018rwt, Harada:2024trx}.

Current cosmological observations favour plateau-like models for the
semi-classical regi\-me of inflation. When extrapolated outside the
observed cosmological window, plateau-models asymptote to a flat and
unbounded potential shape in which stochastic inflation has to take
place. As can be seen in \cref{eq:langevin}, when $V(\phi)$ is exactly
constant, the coarse-grained field is only driven by the noise term
and one can show that $\ev{\calN} = \infty$. The stochastic curvature
fluctuation $\zetafw= \calN - \ev{\calN}$ becomes ill-defined and no
classical trajectory exists to set a privileged $\Nflat$. This
reflects the fact that the system is being driven by pure quantum
effects.

In Ref.~\cite{Blachier:2025tcq}, we have proposed a time-reversed
approach to deal with these exactly flat and unbounded potentials. The
main idea is to solve the stochastic inflation backward, from the end
of the quantum diffusion regime to the initial state, i.e., imposing
from the very beginning that observers have to be in the regions of
the universe where inflation ended. As detailed in
Ref.~\cite{Blachier:2025tcq}, time-reversing stochastic inflation
happens to be equivalent to condition, and then to marginalise, the
stochastic realisations of \cref{eq:langevin} by their lifetimes
$\rv{\Nzero}$, the realisations of $\calN$. Combined with the $\delta
N$-formalism, this method has allowed us to derive, for the first
time, an exact form for the probability distribution
$\Pof{\zeta|\phizero}$ in semi-infinite flat potentials. It is
normalisable with power-law tails $\Pof{\zeta|\phizero} \propto
|\zeta|^{-3/2}$ and does not have any finite moments. This latter
property justifies, a posteriori, the problems of the forward
formalism in which $\ev{\calN}=\infty$. Let us remark that
conditioning the stochastic realisations by $\rv{\Nzero}$ avoids
specifying the, here ill-defined, reference {\efold} number $\Nflat$
to measure $\zeta$. Instead, curvature fluctuations are calculated
over all realisations producing the same lifetime, for all
lifetimes. If we allow ourselves to link the time-reversal approach to
the underlying quantum processes from which \cref{eq:langevin} is
derived, we are somehow enforcing that curvature fluctuations are only
measured once the quantum states have collapsed into an actual amount
of expansion given by $\rv{\Nzero}$. In the absence of any classical
background, this is quite satisfying as local observers have to be
defined by some quantum observable and a measurement: the total amount
of accelerated expansion. In other words, the observed background is
now defined as one of its possible quantum realisations. Let us stress
that, in the exactly flat and unbounded potentials, the time-reverse
approach encompasses all stochastic realisations, including the ones
in which inflation ends after an infinite amount of e-folds.

In this work, we solve time-reversed stochastic inflation for a flat
semi-infinite potential in presence of a constant friction. This
corresponds to a non-vanishing first term on the right-hand side of
\cref{eq:langevin}, which can be viewed as considering a small linear
tilt of the potential. As shown below, the constant friction, or
drift, term does not alter, at all, the time-reversed quantum
diffusion but regularises the probability distribution of the
lifetimes: $\ev{\calN}<\infty$. Therefore, even if the system does not
really have a classical background, a non-vanishing friction permits a
comparison of the probability distribution for $\zeta$ obtained
between the forward and time-reversed formalism.

The paper is organised as follow. In \cref{sec:trsi}, we briefly recap
the time-reversed stochastic inflation formalism, as developed in
Ref.~\cite{Blachier:2025tcq}. The time-reversed equations, in presence
of a constant drift, are then derived and exactly solved. In
\cref{sec:pzeta}, we calculate the probability distribution of the
curvature fluctuations at given lifetime, and marginalised over all
lifetimes. In \cref{sec:discuss}, we quantitatively compare the
probability distributions for $\zeta$ obtained in the time-reversed
and forward approaches. Although they exhibit the same qualitative
behaviour, exponential decaying tails multiplied by power-law, they do
differ by order unity factors in the exponents. Other statistical
properties, such as the mode and the asymmetry between positive and
negative curvatures, are found to be always different. We finally
discuss the extreme case of a negative drift for which eternal
inflation occurs. The differences between the two approaches are
exacerbated, the reverse picture predicting finite curvature
distributions whereas the forward one ends up being pathological. We
explain these results and draw our conclusion in \cref{sec:conc}.

\section{Time-reversed stochastic inflation}
\label{sec:trsi}

\subsection{General formalism}
\label{sec:trform}

The coarse-grained field evolution of equation~\eqref{eq:langevin} is an
It\^o stochastic differential equation of the form
\begin{equation}
\dd{\phi} = F[\phi(N),N]\dd{N} + G[\phi(N),N] \dd{W},
\label{eq:itofw}
\end{equation}
where $\dd W=\xi(N) \dd{N}$ is a Wiener process, $F=-V_{,\phi}/(3H^2)$
is the so-called drift term and $G=H/(2\pi)$ the diffusion
amplitude. The transition probability distribution associated with
\cref{eq:itofw} is solution of the Fokker-Planck (or forward
Kolmogorov) equation given by~\cite{Sarkka_Solin_2019}
\begin{equation}
\pdv{\Pfw{\phi,N|\phizero,\Nzero}}{N} =
-\pdv{}{\phi}\left[F(\phi,N)\Pfw{\phi,N|\phizero,\Nzero}\right] +
\dfrac{1}{2}\pdv[2]{}{\phi}\left[G^2(\phi,N) \Pfw{\phi,N|\phizero,\Nzero}\right],
\label{eq:fp}
\end{equation}
where $N>\Nzero$ and $\Pfw{\phi,\Nzero|\phizero,\Nzero} =
\delta(\phi-\phizero)$. Solving this equation with adequate boundary
conditions gives the probability of finding field values $\phi$ at the
time $N$ starting from the state $\phizero$ at time $\Nzero$. In the
semi-infinite potential, if we denote by $\phiqw$ the field value at
which quantum diffusion ends ($\phiqw \le \phizero$), it corresponds
to an absorbing boundary condition $\Pfw{\phiqw,N|\phizero,\Nzero}=0$,
referred to as the ``quantum wall'' in Ref.~\cite{Blachier:2025tcq}.

Time-reversed stochastic inflation describes the evolution of the same
process but with $\phi$ emerging at $\phi=\phiqw$ and randomly
evolving towards a sink located at $\phi=\phizero$. The starting times
have to be the ending times of the forward processes, i.e., the first
passage times $\Nqw$ at which each realisation of \cref{eq:itofw}
reaches the absorbing boundary at $\phi=\phiqw$. These random times
are in one-to-one mapping with the lifetimes of the reverse
processes. Denoting by $\Plt{\Nqw|\phizero,\Nzero}$ the probability
distribution of the $\Nqw$, the survival probability $S(N)$ at time
$N$ reads~\cite{Ando:2020fjm}
\begin{equation}
S(N|\phizero,\Nzero) =
\int_{\phiqw}^{+\infty}\Pfw{\phi,N|\phizero,\Nzero}\dd{\phi} = 1 -
\int_{\Nzero}^{N}\Plt{\Nqw|\phizero,\Nzero} \dd{\Nqw}.
\label{eq:survival}
\end{equation}
Differentiating by $N$ and using \cref{eq:fp} gives
\begin{equation}
\Plt{N|\phizero,\Nzero} = -\pdv{S(N|\phizero,\Nzero)}{N} =
\dfrac{1}{2}
\eval{\pdv{\left[G^2(\phi,N)\Pfw{\phi,N|\phizero,\Nzero} \right]}{\phi}}_{\phiqw},
\label{eq:Plt}
\end{equation}
where regularity has been assumed at infinity. The time-reversed number of
{\efolds} is defined by
\begin{equation}
\rv{N} \equiv \Nqw - N,
\label{eq:rvN}
\end{equation}
ranging from $\rv{N}=0$, when the field is on the quantum wall, to
$\rv{\Nzero}=\Nqw-\Nzero$ when it reaches the sink at $\phizero$. Let
us stress that $\calN$ of the forward formalism and
$\rv{\Nzero}=\Nqw-\Nzero$ are the same quantities, the probability
distribution of the lifetimes and first passage times are identical
and given by \cref{eq:Plt}.

As shown in Refs.~\cite{Nagasawa1964,Anderson1982,Blachier:2025tcq},
the time-reversed process follows an It\^o stochastic equation which
is completely characterised by the probability distribution
$\Prv{\phi,\rv{N}|\phizero,\rv{\Nzero}}$, with $\rv{N} \le
\rv{\Nzero}$, solution of the time-reversed Fokker-Planck equation
\begin{equation}
  \begin{aligned}
\pdv{\Prv{\phi,\rv{N}|\phizero,\rv{\Nzero}}}{\rv{N}} & =
-\pdv{\phi}\left[\Frv(\phi,\rv{N})
  \Prv{\phi,\rv{N}|\phizero,\rv{\Nzero}}\right] \\ & + \dfrac{1}{2}
\pdv[2]{}{\phi}\left[G^2(\phi,N)\Prv{\phi,\rv{N}|\phizero,\rv{\Nzero}}\right],
\end{aligned}
\label{eq:rvfp}
\end{equation}
where the reversed drift term $\Frv(\phi,\rv{N})$ reads
\begin{equation}
    \Frv(\phi,\rv{N}) = -F(\phi,N) +
    \dfrac{1}{\Pfw{\phi,N|\phizero,\Nzero}} \pdv{}{\phi}\left[G^2(\phi,N)\Pfw{\phi,N|\phizero,\Nzero} \right],
\label{eq:rvdrift}
\end{equation}
the diffusion coefficient $G$ being unchanged.
Notice the appearance of the forward transition probability
$\Pfw{\phi,N|\phizero,\Nzero}$ in this expression, requiring to first
solve \cref{eq:fp} before moving on determining
$\Prv{\phi,\rv{N}|\phizero,\rv{\Nzero}}$ and thereby rendering the
time-reversed approach technically more involved.

\subsection{Quantum diffusion with friction}
\label{sec:fric}

For a self-gravitating scalar field, the first Friedmann-Lema\^{\i}tre equation can
be recast into~\cite{Martin:2016iqo, Auclair:2024udj}
\begin{equation}
H^2 = \dfrac{V(\phi)}{3 - \eps{1}}\,,
\end{equation}
where the first Hubble flow function maps the field kinetic energy
\begin{equation}
\eps{1} = \dfrac{1}{2} \left(\dv{\phi}{N}\right)^2.
\end{equation}
For an exactly flat semi-infinite potential at $\phi \ge \phiqw$, one
has $V=\Vzero$ and the drift term $F$ in \cref{eq:itofw} vanishes. The
stochastic field $\phi$ is then driven by pure quantum diffusion with
$G=H/(2\pi)$ and, provided $\Vzero \ll 1$ (sub-Planckian values), one has
$\eps{1} \ll 1$ and $H^2 \simeq \Hinf^2=\Vzero/3$ is
constant. Time-reversed stochastic inflation for this case has been
exactly solved in Ref.~\cite{Blachier:2025tcq}.

In order to tame the effects associated with pure quantum diffusion,
we now consider a non-vanishing drift term $F$ in
\cref{eq:itofw}. Such a term can appear if, instead of being exactly
flat, the potential is slightly tilted at $\phi > \phiqw$. For
instance, considering
\begin{equation}
V(\phi) = \Vzero\left(1 + \alpha \dfrac{\rv{\phi}}{\rv{\phiinf}} \right),
\label{eq:tiltpot}
\end{equation}
where $\alpha$ is a dimensionless number, $\rv{\phi}\equiv
\phi-\phiqw$ and $\rv{\phiinf} = \phiinf-\phiqw$ is a (large) field
excursion domain, one has
\begin{equation}
\dfrac{V_{,\phi}}{3 H^2} \simeq \dfrac{V_{,\phi}}{V} \simeq \dfrac{\alpha}{\rv{\phiinf}}\,,
\end{equation}
provided $\rv{\phi} \ll \rv{\phiinf}/\alpha$. The previous
approximation assumes that $2 \eps{1} \simeq \alpha^2/\rv{\phiinf}^2
\ll 1$, which also demands that $\rv{\phiinf}/\alpha \gg 1$ (Planck
units). When both conditions are satisfied, $V(\phi) \simeq \Vzero$
and $H^2 \simeq \Hinf^2 = \Vzero/3$ remains constant. To summarize, if
$\alpha$ is of order unity, for a very large excursion domain
$\rv{\phiinf}$, and provided the stochastic field remains such that
$\phi-\phiqw < \rv{\phiinf}$, stochastic inflation is described by a
Langevin equation with constant friction and diffusion terms
\begin{equation}
\dd{\phi} = -\dfrac{\alpha}{\rv{\phiinf}} \dd{N} + \dfrac{\Hinf}{2\pi}
\xi \dd{N}.
\label{eq:langedrift}
\end{equation}
This is the same equation as for the so-called ``tilted quantum well''
discussed in the context of PBH in Refs.~\cite{Ezquiaga:2019ftu,
  Animali:2022otk, Animali:2024jiz}. However, because the quantum
diffusion domain here is semi-infinite, as opposed to bounded,
\cref{eq:langedrift} breaks down if the stochastic field develops
excursions larger than $\rv{\phiinf}$. Indeed, neither $H(\phi)$ nor
$V(\phi)$ can be considered constant in these cases. Nonetheless, and for
reasons that will become clear later on, such a situation is under
control in the time-reversed formalism as it ends up being dependent of
the lifetimes. Moreover, it is always possible to take very large
$\rv{\phiinf}$ values, or, to transit to an exactly flat potential at
$\rv{\phi} > \rv{\phiinf}/\alpha$.

From now on, we will be interested in solving \cref{eq:fp,eq:rvfp}
with\footnote{Notice that the friction term has the dimension of a
mass, it reads $F=-\alpha \Mp^2/\rv{\phiinf}$ with the Planck mass explicit.}
\begin{equation}
F = -\dfrac{\alpha}{\rv{\phiinf}}\,, \qquad G = \dfrac{\Hinf}{2\pi}\,.
\label{eq:FandG}
\end{equation}

\subsection{Forward transition probability}

The forward transition probability $\Pfw{\phi,N|\phizero,\Nzero}$ is
the solution of \cref{eq:fp} submitted to the absorbing boundary and initial
conditions
\begin{equation}
  \Pfw{\phiqw,N|\phizero,\Nzero}=0, \qquad
  \Pfw{\phi,\Nzero|\phizero,\Nzero} = \dirac{\phi-\phizero}.
\end{equation}
It can be obtained by a Fourier transform, using the superposition
principle, or the method of images, to enforce the absorbing
condition. One obtains, for $N \ge \Nzero$ and $\phi \ge \phiqw$,
\begin{equation}
  \begin{aligned}
\Pfw{\phi,N|\phizero,\Nzero} = \dfrac{1}{\sqrt{2\pi} G
  \sqrt{N-\Nzero}} \left[ e^{-\frac{\left(\phi-\phizero +\alpha
      \frac{N-\Nzero}{\rv{\phiinf}} \right)^2}{2 G^2
    \left(N-\Nzero\right)}} \right. & - \left.  e^{-\frac{2\alpha \left(\phiqw -
    \phizero\right)}{G^2 \rv{\phiinf}}} \right. \\ & \times \left.
e^{-\frac{\left(\phi-2\phiqw+\phizero+ \alpha \frac{N-\Nzero}{\rv{\phiinf}} \right)^2}{2
        G^2 \left(N-\Nzero\right)}} \right].
\end{aligned}
\label{eq:Pfwsol}
\end{equation}
The probability distribution of the first-passage times $\Nqw$ at
$\phi = \phiqw$ is given by \cref{eq:Plt}. This is also the
probability distribution of the lifetimes $\rv{\Nzero}=\Nqw-\Nzero$
and one gets
\begin{equation}
\Plt{\rv{\Nzero}|\phizero} = \Plt{\Nqw|\phizero,\Nzero} =
\dfrac{\rv{\phizero}}{\sqrt{2\pi}\,G\,\rv{\Nzero}^{3/2}}\,
e^{-\frac{\left(\rv{\phizero} - \alpha \frac{\rv{\Nzero}}{\rv{\phiinf}} \right)^2}{2 G^2 \rv{\Nzero}}},
\label{eq:Pltsto}
\end{equation}
where the initial field value is now in reference to the quantum wall
\begin{equation}
\rv{\phizero} \equiv \phizero - \phiqw.
\end{equation}
Taking $\alpha=0$ in \cref{eq:Pltsto}, one recovers the distribution
of the lifetimes for the exactly flat and semi-infinite
potential~\cite{Blachier:2025tcq}. In particular, the large lifetime
limit for $\alpha=0$ gives $\Plt{\rv{\Nzero}|\phizero} \propto
\rv{\Nzero}^{-3/2}$ and one recovers that $\ev{\calN} =
\ev{\rv{\Nzero}} = \infty$. On the contrary, for $\alpha > 0$, one
gets an exponential behaviour at infinity ensuring the existence of
all moments.

\subsection{Time-reversal}
\label{sec:revsto}

Applying the time-reversal formalism detailed in \cref{sec:trform},
one needs to solve \cref{eq:rvfp} with a reverse drift term given by
\cref{eq:rvdrift}. From the constancy of \cref{eq:FandG}, one gets
\begin{equation}
\Frv = \dfrac{\alpha}{\rv{\phiinf}} + G^2 \pdv{}{\phi}\left[\ln \Pfw{\phi,N|\phizero,\Nzero}\right],
\label{eq:Frvsto}
\end{equation}
where the forward transition probability distribution is given in
\cref{eq:Pfwsol}. This one can be more conveniently expressed in terms
of the time-reversed {\efold} numbers $\rv{N}$ and the field values in
reference to the quantum wall
\begin{equation}
\rv{\phi} \equiv \phi - \phiqw,
\end{equation}
as
\begin{equation}
\Pfw{\phi,N|\phizero,\Nzero} = \dfrac{ e^{-\frac{\alpha^2\left(\rv{\Nzero}-\rv{N}\right)^2 + 2\alpha
    \rv{\phiinf}\left(\rv{\phi}-\rv{\phizero}\right)}{2
    G^2\rv{\phiinf}^2}} }{\sqrt{2\pi}\,G \sqrt{\rv{\Nzero}-\rv{N}}} \left[e^{-\frac{\left(\rv{\phi}-\rv{\phizero}
      \right)^2}{2G^2\left(\rv{\Nzero}-\rv{N}\right)}} - e^{-\frac{\left(\rv{\phi}+\rv{\phizero}
      \right)^2}{2G^2\left(\rv{\Nzero}-\rv{N}\right)}} \right].
\label{eq:Pfwnice}
\end{equation}
Plugging \cref{eq:Pfwnice} into \cref{eq:Frvsto}, one obtains, after
some algebra,
\begin{equation}
  \begin{aligned}
\Frv(\phi,\rv{N}) & =
-\dfrac{\rv{\phi}-\rv{\phizero}}{\rv{\Nzero}-\rv{N}} +
\dfrac{2\rv{\phizero}}{{\rv{\Nzero}-\rv{N}}} \dfrac{e^{-\frac{2
      \rv{\phizero} \rv{\phi}}{G^2
      \left(\rv{\Nzero}-\rv{N}\right)}}}{1 - e^{-\frac{2 \rv{\phizero}
      \rv{\phi}}{G^2\left(\rv{\Nzero} -\rv{N}\right)}}} \\ & =
-\dfrac{\rv{\phi}}{\rv{\Nzero} -\rv{N}} +
\dfrac{\rv{\phizero}}{\rv{\Nzero} - \rv{N}}
\coth{\left[\dfrac{\rv{\phizero}\rv{\phi}}{G^2\left(\rv{\Nzero}-\rv{N}\right)}\right]}.
\end{aligned}
\label{eq:Frvfric}
\end{equation}
There is no longer any term depending on $\alpha$ in this expression
showing that the reverse drift $\Frv$ is independent of the forward
(constant) drift. As a matter of fact, \cref{eq:Frvfric} is the same
as the one derived in Ref.~\cite{Blachier:2025tcq} [see Eq.~(3.12)]
for the flat semi-infinite potential. As a consequence, the solution
of the time-reversed Fokker-Planck equation~\eqref{eq:rvfp},
satisfying the initial and boundary conditions
\begin{equation}
\Prv{\phi,\rv{N}=0|\phizero,\rv{\Nzero}} = \dirac{\rv{\phi}},\quad
\Prv{\phi,\rv{N}=\rv{\Nzero}|\phizero,\rv{\Nzero}} = \dirac{\rv{\phi}-\rv{\phizero}},
\label{eq:Prvibc}
\end{equation}
is \emph{exactly} the same as the one derived for the flat
potential. Introducing the rescaled variables
\begin{equation}
\chihat \equiv \dfrac{\rv{\phi}}{G\sqrt{\rv{\Nzero}}}\,, \qquad \chizerohat \equiv
\dfrac{\rv{\phizero}}{G\sqrt{\rv{\Nzero}}}\,,\qquad \tau \equiv
\dfrac{\rv{N}}{\rv{\Nzero}},
\end{equation}
it reads~\cite{Mazzolo2024, Blachier:2025tcq}
\begin{equation}
\Prv{\phi,\rv{N}|\phizero,\rv{\Nzero}}  =
\dfrac{\sqrt{2/\pi}}{\tau^{3/2} \sqrt{1-\tau}} \,
\dfrac{\chihat}{\chizerohat G \sqrt{\rv{\Nzero}}}
\sinh\left(\dfrac{\chihat\chizerohat}{1-\tau}\right)
e^{-\frac{\chihat^2 + \tau^2 \chizerohat^2}{2 \tau(1-\tau)}}.
\label{eq:Prvnice}
\end{equation}
\begin{figure}
\begin{center}
  \includegraphics*[width=\onefigw]{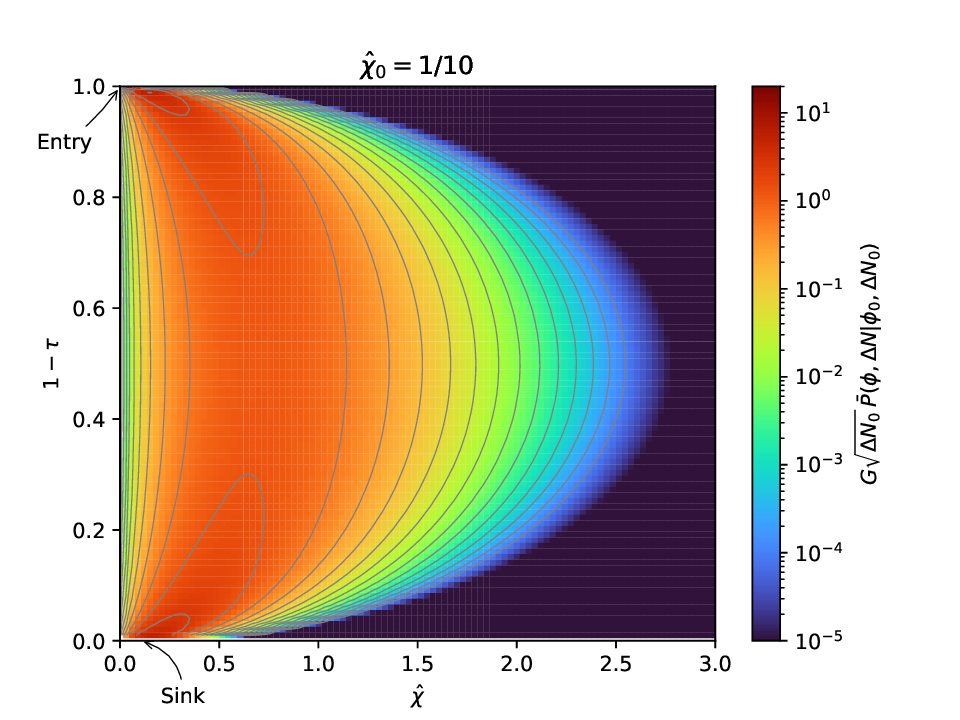}
  \includegraphics*[width=\onefigw]{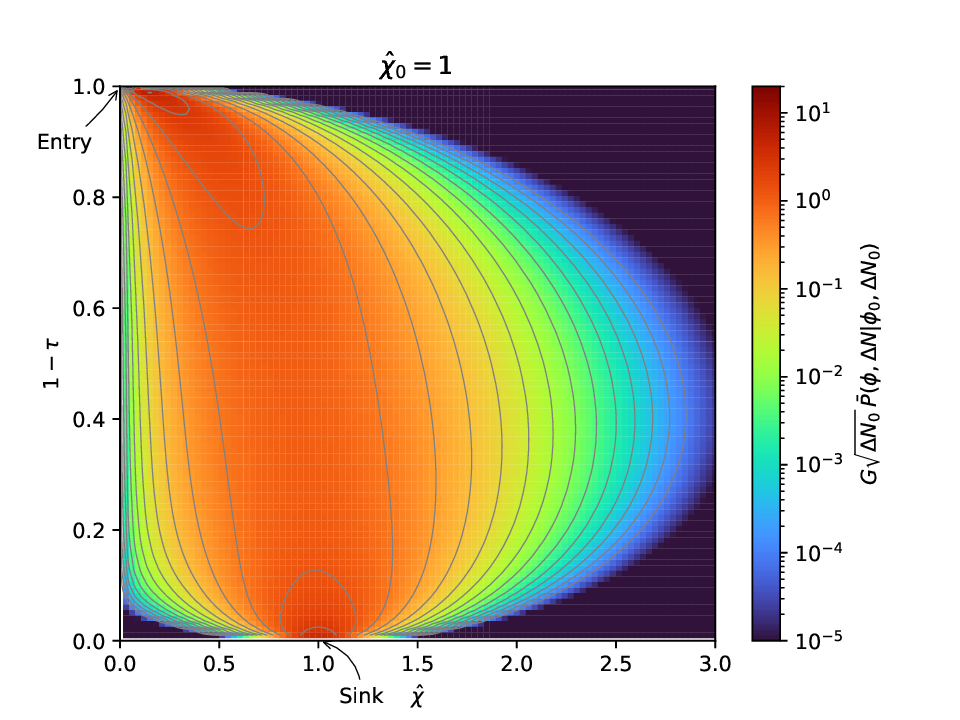}
\caption{Contour plots of the rescaled reverse probability
  distribution $G \sqrt{\rv{\Nzero}} \,
  \Prv{\phi,\rv{N}|\phizero,\rv{\Nzero}}$ as a function of $\chihat =
  \rv{\phi}/(G\sqrt{\rv{\Nzero}})$ and $\tau=\rv{N}/\rv{\Nzero}$ in
  the diffusion regime ($\chizerohat \le 1$). The field emerges at
  $\chihat=0$ at time $\tau=0$, diffuses in the domain before being
  absorbed at $\chihat=\chizerohat$ at $\tau=1$.}
\label{fig:Prvdiff}
\end{center}
\end{figure}

\begin{figure}
\begin{center}
  \includegraphics*[width=\onefigw]{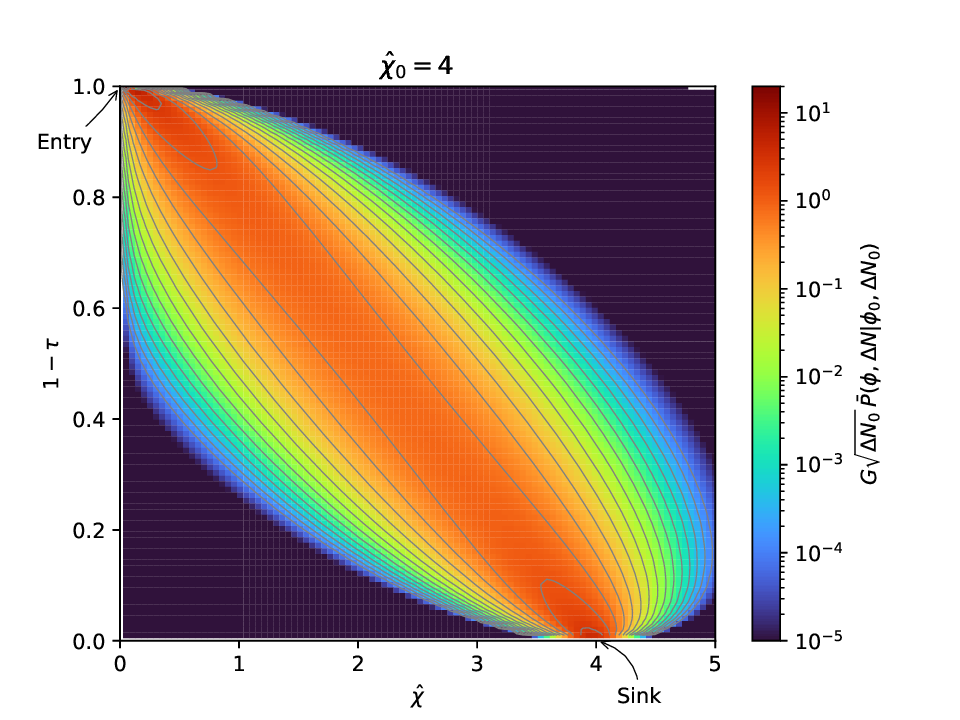}
  \includegraphics*[width=\onefigw]{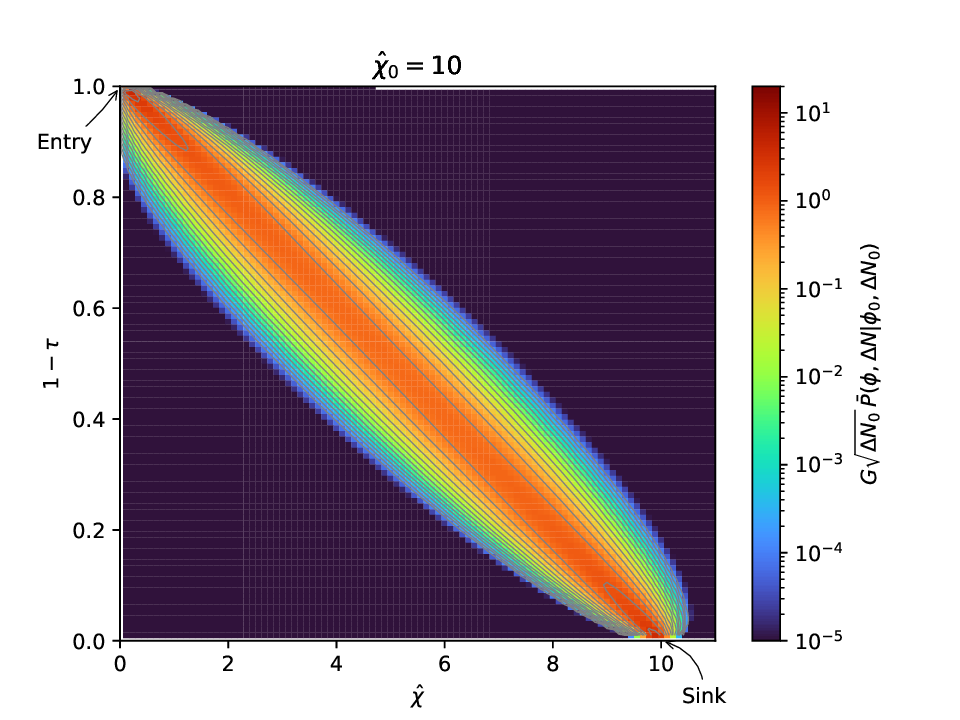}
\caption{Contour plots of the rescaled reverse probability
  distribution $G \sqrt{\rv{\Nzero}} \,
  \Prv{\phi,\rv{N}|\phizero,\rv{\Nzero}}$ as a function of $\chihat =
  \rv{\phi}/(G\sqrt{\rv{\Nzero}})$ and $\tau=\rv{N}/\rv{\Nzero}$ in
  the fluxing regime ($\chizerohat > 1$). The field emerges at
  $\chihat=0$ at time $\tau=0$ and straightly jumps to
  $\chihat=\chizerohat$ at $\tau=1$.}
\label{fig:Prvflux}
\end{center}
\end{figure}

The new time variable $\tau \in [0,1]$ is the reverse {\efold} number
in unit of the lifetime $\rv{\Nzero}$. Similarly, $\chihat$ are
field values, in reference to the quantum wall, in unit of the typical
diffusion excursion domain, $G \sqrt{\rv{\Nzero}}$, where the
diffusion coefficient $G$ is given by the Hubble parameter during
inflation. As described at length in Ref.~\cite{Blachier:2025tcq},
\cref{eq:Prvnice} completely characterises time-reversed stochastic
inflation: the field emerges at the quantum wall $\phiqw$, at time
$\rv{N}=0$, and randomly evolves to reach the sink $\phizero$ at a
reverse {\efold} time $\rv{N}$ matching the lifetime $\rv{\Nzero}$ of
the forward process. As explicit in \cref{eq:Prvnice}, the
conditioning is with respect to $\phizero$ \emph{and} to the lifetime
$\rv{\Nzero}$. In other words, we have now sorted, and partitioned,
all forward realisations in which stochastic inflation ends by
$\rv{\Nzero}$. Let us remark that it should not come as a surprise
that the constant drift we are considering here does not impact the
reverse process. Indeed, conditioning a Brownian motion, or a drifted
Brownian motion, on its final state is known to lead to the same
process: the so-called Brownian bridge, which is thus independent of
the original drift~\cite{Mazzolo2024}.

The rescaled reverse probability distribution $G \sqrt{\rv{\Nzero}}
\Prv{\phi,\rv{N}|\phizero,\rv{\Nzero}}$ has been represented in
\cref{fig:Prvdiff} in the diffusion dominated regime ($\chizerohat\le
1$) and in \cref{fig:Prvflux} for the fluxing regime ($\chizerohat >
1$). In the diffusion dominated case, $\chizerohat < 1$ implies that
the sink $\phizero$ is located well within the typical excursion
domain $G \sqrt{\rv{\Nzero}}$ associated with quantum diffusion
lasting $\rv{\Nzero}$ {\efolds}. Therefore, the reverse process has
plenty of time to transit from the quantum wall to the sink and goes
on visiting the surrounding with the extra time. On the contrary, for
$\chizerohat \gg 1$, one has $\rv{\phizero} \gg G \sqrt{\rv{\Nzero}}$
and solely the straightest stochastic realisations are selected:
quantum diffusion is much reduced, this is the fluxing regime.

\subsection{To drift or not to drift}
\label{sec:driftdiscuss}

As discussed in \cref{sec:fric}, the Langevin
equation~\eqref{eq:langedrift} with constant drift can be considered
as a good approximation of the one stemming from the tilted potential
of \cref{eq:tiltpot} but only within the domain $\rv{\phi} \ll
\rv{\phiinf}/\alpha$ (provided that $\rv{\phiinf}/\alpha \gg 1$). As
can be seen in \cref{fig:Prvdiff,fig:Prvflux}, this is completely
under control and fixed by the value of $\chizerohat$, i.e., by both
$\phizero$ and $\rv{\Nzero}$. Moreover, because the time-reversed
process is the same as the one with $\alpha=0$, one might assume
that patching the tilted potential to an exactly flat semi-infinite
plateau at $\rv{\phi}=\rv{\phiinf}/\alpha$ would still give the same
reverse probability distribution
$\Prv{\phi,\rv{N}|\phizero,\rv{\Nzero}}$ as in \cref{eq:Prvnice}. This
is not entirely correct as sudden transitions in the potential are
known to trigger sub-Hubble quantum fluctuations different than the
ones required for having a Gaussian white noise
$\xi(N)$~\cite{Jackson:2023obv,Briaud:2025ayt}. Nonetheless, for a
smooth transition between the tilted potential and the exactly flat one at
$\rv{\phi}=\rv{\phiinf}/\alpha$, it is reasonable to assume that
$\Prv{\phi,\rv{N}|\phizero,\rv{\Nzero}}$ will be only slightly
distorted in that region. In other words, the constant friction we are
considering, and thus the linear tilt of the potential, has
essentially only one effect which is to change the probability
distribution of the lifetimes without affecting the strong quantum
randomness of the system. This intuitively implies that considering a
small tilt or no tilt at all does not drastically change the quantum
dynamics in the semi-infinite potentials.

In the next section, we derive the probability distribution of the
curvature fluctuations using the reverse-time stochastic $\delta
N$-formalism and show that, up to the tails, it is indeed similar to
one stemming from the flat semi-infinite potential.

\section{Quantum-generated curvature distribution}
\label{sec:pzeta}

In order to extract the curvature distribution, we apply the adapted
version of the stochastic $\delta N$-formalism to the time-reversed
process, as exposed in Ref.~\cite{Blachier:2025tcq}. The reference
$\Nflat$ is taken as the stochastic average within the reverse processes,
and $\zeta = N-\Nzero -\ev{N-\Nzero}$ is promoted to a stochastic
variable conditioned by the lifetime $\rv{\Nzero}$, i.e.,
\begin{equation}
\zeta = \ev{\rv{N}} - \rv{N}.
\label{eq:trzeta}
\end{equation}

As represented in \cref{fig:Prvdiff,fig:Prvflux}, at given lifetime
$\rv{\Nzero}$, quantum diffusion allows for many different
realisations and, for a given field value $\phi$, $\rv{N}$ is a
stochastic quantity. As explicit in \cref{eq:trzeta}, $\zeta$ is
measured by the fluctuations of the reverse number of {\efolds}
around the mean, in a background defined by the lifetime $\rv{\Nzero}$
of the processes.

\subsection{Curvature fluctuations at given lifetime}
\label{sec:zetagivenLT}

The reverse probability distribution
$\Prv{\phi,\rv{N}|\phizero,\rv{\Nzero}}$ being identical to the one
derived for the flat semi-infinite potential, the calculations
required to determine the curvature fluctuations, at given $\phizero$
and $\rv{\Nzero}$, are the same as in Ref.~\cite{Blachier:2025tcq}. As
such, we only summarise below the main steps on how to obtain the
probability distribution $\Pof{\zeta|\phizero,\rv{\Nzero}}$.

As visible in \cref{fig:Prvdiff,fig:Prvflux}, a given field value can
be reached by many trajectories thereby promoting $\rv{N}$ to a random
variable. Therefore, one has first to determine
$\Pof{\rv{N}|\phi,\phizero,\rv{\Nzero}}$, which is given by
$\Prv{\phi,\rv{N}|\phizero,\rv{\Nzero}}$ up to a normalisation factor
which has to be calculated. This probability distribution allows us to
calculate $\ev{\rv{N}} = \ev{\tau} \rv{\Nzero}$ and one
gets~\cite{Blachier:2025tcq}
\begin{equation}
\ev{\tau} =  \rv{\Nzero} \int_0^1
\tau \, \Pof{\tau\rv{\Nzero}|\phi,\phizero,\rv{\Nzero}}
\dd{\tau} =  \sqrt{\dfrac{\pi}{2}} \,
  \chihat \, e^{\frac{\chizerohat^2}{2}} \, \dfrac{\erf\left(\dfrac{2
      \chihat + \chizerohat}{\sqrt{2}}\right) -
    \erf\left(\dfrac{\chihat +
      \left|\chihat-\chizerohat\right|}{\sqrt{2}}\right)}{e^{-\chihat\left(\left|\chihat-\chizerohat\right|
      + \chihat-\chizerohat\right)} -
    e^{-2\chihat\left(\chihat+\chizerohat\right)}}\,.
\label{eq:taumean}
\end{equation}

From \cref{eq:trzeta}, the joint distribution
$\Pof{\phi,\zeta|\phizero,\rv{\Nzero}}$ is obtained by plugging the
relation $\rv{N} = \rv{\Nzero} \ev{\tau} -\zeta$ in
\cref{eq:Prvnice}. Defining the curvature fluctuation in unit of the
lifetime
\begin{equation}
\zetahat \equiv \dfrac{\zeta}{\rv{\Nzero}}\,,
\label{eq:zetahatdef}
\end{equation}
one obtains
\begin{equation}
  \begin{aligned}
\Pof{\phi,\zetahat|\phizero,\rv{\Nzero}} & = \dfrac{1}{G\sqrt{\rv{\Nzero}}}
\dfrac{1}{\sqrt{2\pi}} \dfrac{\chihat}{\chizerohat}
\dfrac{\heaviside{\ev{\tau}-\zetahat} - \heaviside{\ev{\tau} -
    \zetahat -1}}{\left(\ev{\tau}-\zetahat\right)^{3/2}
  \sqrt{1-\ev{\tau} + \zetahat}}  \\ & \times \exp\left\{-\dfrac{\left[\chihat -
  \left(\ev{\tau} - \zetahat\right) \chizerohat\right]^2}{2
  \left(\ev{\tau} - \zetahat\right)\left(1-\ev{\tau} +
  \zetahat\right)} \right\}\left[1 -
  \exp\left(-\dfrac{2\chihat\chizerohat}{1-\ev{\tau} +
    \zetahat}\right) \right].
\end{aligned}
\label{eq:PzetaphiGivenLT}
\end{equation}
The Heaviside functions appearing in this expression come from the
reverse process enforcing that $0<\rv{N} < \rv{\Nzero}$, which gets
translated into $0< \ev{\tau} - \zetahat < 1$ in the $\delta
N$-formalism. Because $\ev{\tau}$ is a (complicated) function of
$\chihat$, see \cref{eq:taumean}, these functions act as complicated
windows on field values selecting only the domains compatible with the
given curvature.

\begin{figure}
\begin{center}
  \includegraphics*[width=\onefigw]{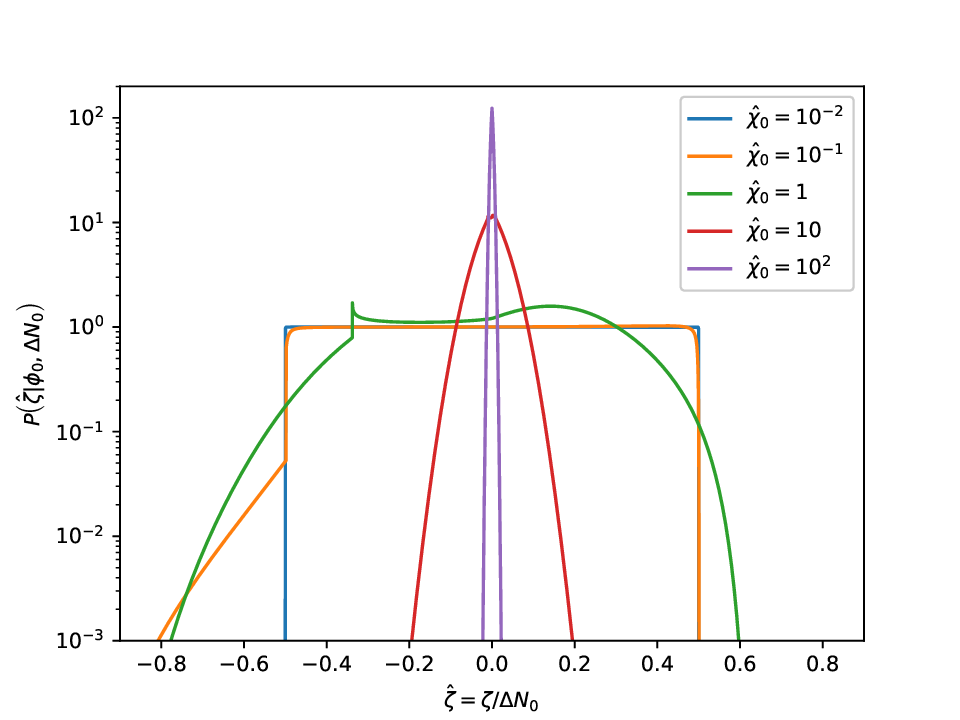}
  \includegraphics*[width=\onefigw]{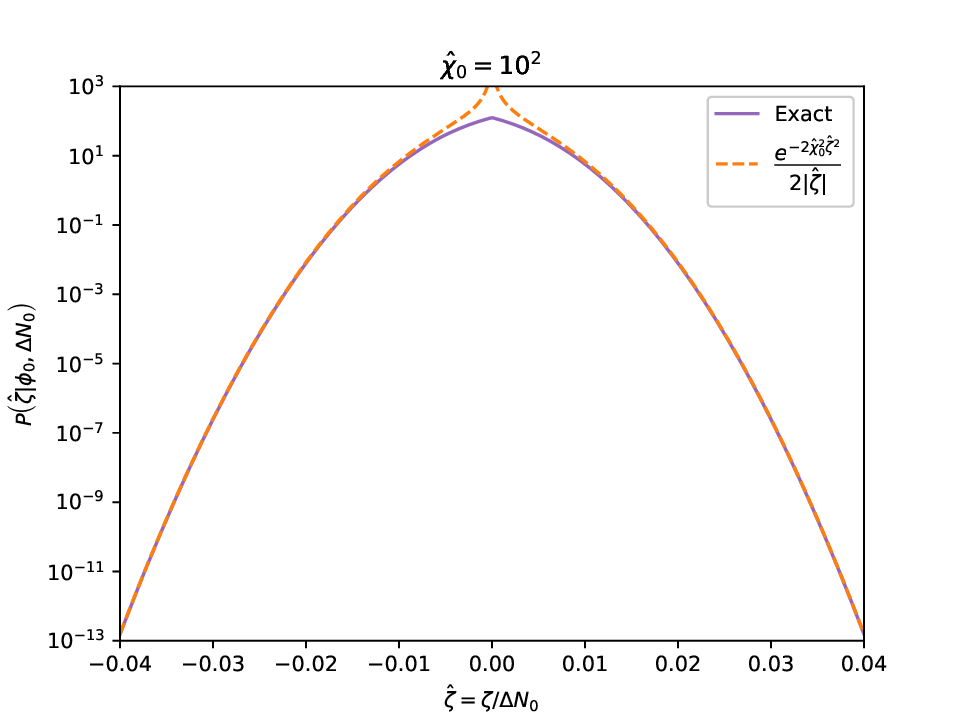}
\caption{Probability distribution of the rescaled curvature
  fluctuations $\zetahat = \zeta/\rv{\Nzero}$, at given lifetime
  $\rv{\Nzero}$, as a function of its sole parameter
  $\chizerohat=\rv{\phizero}/(G\sqrt{\rv{\Nzero}})$ (top panel). It
  asymptotes to the rectangle function of \cref{eq:Pzetatauhalf} in the
  large diffusion limit (small $\chizerohat$), and to a Dirac-like
  distribution in the fluxing regime (large $\chizerohat$). In
  between, it develops a complicated shape, some visible kicks coming
  from the field-domain selection functions appearing in
  \cref{eq:PzetaphiGivenLT}. The bottom panel is a zoom on the
  distribution in the fluxing regime, for $\chizerohat=10^2$. The
  tails of the distribution are becoming Gaussian and decay as
  $e^{-2\chizero^2 \zetahat^2}/(2 |\zetahat|)$.}
\label{fig:Pzetahat}
\end{center}
\end{figure}

Finally, a marginalisation over all field values gives the
probability distribution of $\zetahat$ given $\phizero$ and
$\rv{\Nzero}$
\begin{equation}
\Pof{\zetahat|\phizero,\rv{\Nzero}} = \int_0^{+\infty}
\Pof{\phi,\zetahat|\phizero,\rv{\Nzero}} \dd{\rv{\phi}} = G
\sqrt{\rv{\Nzero}} \int_0^{+\infty} \Pof{\phi,\zetahat|\phizero,\rv{\Nzero}} \dd{\chihat}.
\label{eq:PzetahatGivenLT}
\end{equation}
Let us stress that $\zetahat$ in these expressions is the stochastic
variable. From \cref{eq:zetahatdef}, one has
\begin{equation}
\Pof{\zeta|\phizero,\rv{\Nzero}} = \dfrac{1}{\rv{\Nzero}}
\Pof{\zetahat|\phizero,\rv{\Nzero}}.
\label{eq:PzetaGivenLT}
\end{equation}
Moreover, the functional forms of \cref{eq:taumean,eq:PzetaphiGivenLT}
imply that $\Pof{\zetahat|\phizero,\rv{\Nzero}}$ is a function of
$\chizerohat$ only. The integral over $\chihat$ in
\cref{eq:PzetahatGivenLT} has to be performed numerically and we have
plotted the results in \cref{fig:Pzetahat}. However, in the diffusion
regime displayed in \cref{fig:Prvdiff}, one has $\chizerohat \ll 1$
and \cref{eq:taumean} implies that $\ev{\tau} \to 1/2$. Plugging
$\ev{\tau}=1/2$ in \cref{eq:PzetaphiGivenLT}, the integral of
\cref{eq:PzetahatGivenLT} is explicit and
reads~\cite{Blachier:2025tcq}
\begin{equation}
\Pof{\zetahat|\phizero,\rv{\Nzero}} \underset{\ev{\tau}\to 1/2}{=}
\heaviside{\dfrac{1}{2}-\zetahat} - \heaviside{-\dfrac{1}{2} - \zetahat}.
\label{eq:Pzetatauhalf}
\end{equation}
In the pure diffusion limit, all values of $-1/2<\zetahat<1/2$ are
equiprobable, and so it is for $-\rv{\Nzero}/2 <
\zeta<\rv{\Nzero}/2$. This is confirmed by the numerical results
displayed in \cref{fig:Pzetahat}, but could equally well be spotted by
eyes in the top panel of \cref{fig:Prvdiff}.

As evident in \cref{fig:Pzetahat}, in the fluxing regime, for
$\chizerohat \gg 1$, the curvature fluctuations are much reduced and
the probability distribution for $\zetahat$ approaches a Dirac-like
distribution. In fact, it is possible to show that, in the limit of
large $\chizerohat$, one has $\ev{\tau} \simeq
\chihat/\chizerohat$. Plugging this approximation into
\cref{eq:PzetaphiGivenLT}, the various integrals appearing in
\cref{eq:PzetahatGivenLT} can be analytically performed\footnote{See
\cref{sec:marge} and appendix B of Ref.~\cite{Blachier:2025tcq} on how
to deal with these integrals.} to show that
\begin{equation}
  \lim_{\chizerohat |\zetahat| \to  \infty}
  \Pof{\zetahat|\chizero,\rv{\Nzero}} \simeq \dfrac{e^{-2\chizerohat^2
      \zetahat^2}}{2|\zetahat|}.
\label{eq:Pzetahatfluxing}
\end{equation}
The tails of the curvature distribution at given lifetime are becoming
Gaussian. The approximation of \cref{eq:Pzetahatfluxing} has been
represented in the bottom panel of \cref{fig:Pzetahat}, for
$\chizerohat=100$, together with the distribution obtained from an
exact numerical integration.

\subsection{Marginalisation over lifetimes}
\label{sec:marge}

The results summarised in the previous section are identical to the
one associated with the exactly flat semi-infinite potential. The
friction term modifying only the probability distribution of the
lifetimes, it enters into the marginalisation of
$\Pof{\zeta|\phizero,\rv{\Nzero}}$ over $\rv{\Nzero}$. Explicitly,
using \cref{eq:PzetaGivenLT}, one
has
\begin{equation}
\Pof{\zeta|\phizero} = \int_0^{+\infty}
\Pof{\zetahat|\phizero,\rv{\Nzero}} \dfrac{\Plt{\rv{\Nzero}|\phizero}}{\rv{\Nzero}}
\dd{\rv{\Nzero}},
\label{eq:Pzeta}
\end{equation}
where the distribution of the lifetimes is given in
\cref{eq:Pltsto}. We should remember that
$\Pof{\zetahat|\phizero,\rv{\Nzero}}=\Pof{\zetahat|\chizerohat}$ ends
up being a function of $\chizerohat$ only, which suggests to change
the integration variable from $\rv{\Nzero}$ to $\chizerohat =
\rv{\phizero}/(G \sqrt{\rv{\Nzero}})$ (at fixed $\rv{\phizero}$). One has to
pay attention to the argument of this function, which is
\begin{equation}
\zetahat = \dfrac{\zeta}{\rv{\Nzero}} = \dfrac{\zeta}{\chizero^2}
\times \chizerohat^2,  
\end{equation}
where we have defined
\begin{equation}
\chizero \equiv \dfrac{\rv{\phizero}}{G}\,,
\end{equation}
the initial field value, in reference to the quantum wall, in unit of
the quantum diffusion $G$. After some algebra, \cref{eq:Pzeta}
reads\footnote{The upper bound comes from the window functions of
\cref{eq:PzetaphiGivenLT} enforcing $0< \ev{\tau} - \zetahat < 1$ with
$\ev{\tau}\in[0,1]$.}
\begin{equation}
\Pof{\zeta|\phizero} = \dfrac{e^{\alphahat}}{\chizero^2} \sqrt{\dfrac{2}{\pi}}
\int_0^{\frac{\chizero}{\sqrt{\abs{\zeta}}}}
\Pof{\left.\dfrac{\zeta}{\chizero^2}\chizerohat^2\right|\chizerohat} \chizerohat^2
e^{-\frac{\chizerohat^2}{2} - \frac{\alphahat^2}{2\chizerohat^2}} \dd{\chizerohat},
\label{eq:Pzetanice}
\end{equation}
in which we have introduced the rescaled drift
coefficient\footnote{Notice that $\alphahat$ is dimensionless,
restoring the Planck mass, it indeed reads $\alphahat =
\dfrac{\alpha \Mp^2}{G^2}\dfrac{\rv{\phizero}}{\rv{\phiinf}} $.}
\begin{equation}
\alphahat \equiv
\dfrac{\alpha}{G^2}\dfrac{\rv{\phizero}}{\rv{\phiinf}} =
\dfrac{\alpha}{G} \dfrac{\chizero}{\rv{\phiinf}}\,.
\label{eq:alphahatdef}
\end{equation}
Without performing any calculations, it is clear that, up to the 
$e^{\alphahat}$ normalisation factor, the friction term $\alphahat$
acts as a regulator for $\chizerohat$ small (diffusion regime) while
having minimal impact for $\chizerohat \gg 1$ (fluxing regime). Moreover,
\cref{eq:Pzetanice} shows that $\Pof{\zeta|\phizero}$ is a function of
$\alphahat$ and $\zeta/\chizero^2$ only.

\begin{figure}
\begin{center}
  \includegraphics*[width=\onefigw]{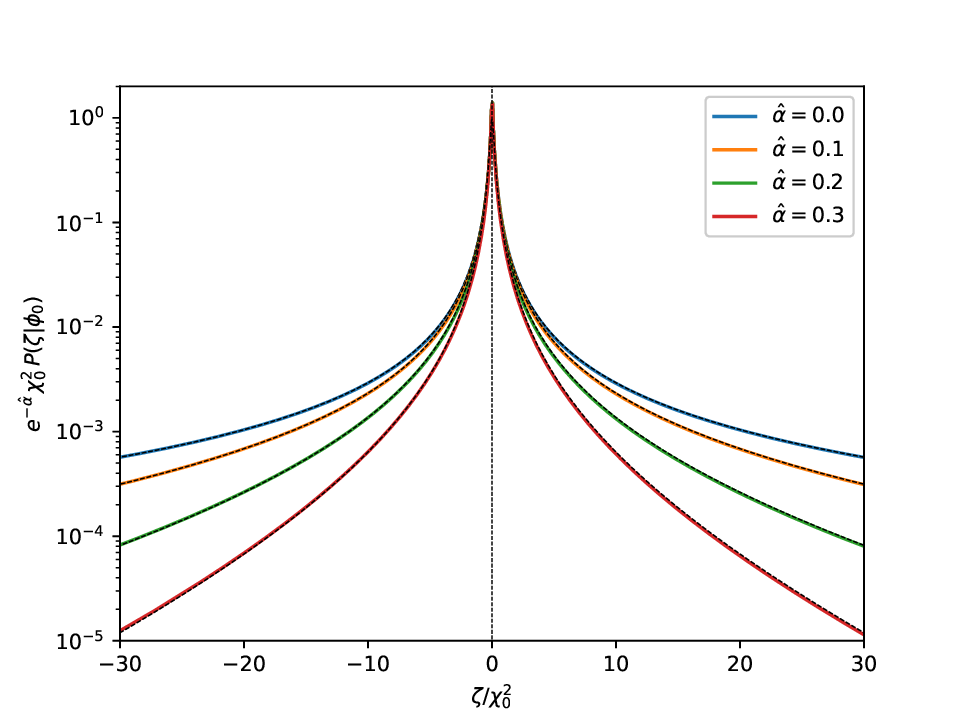}
\caption{Probability distribution of the curvature fluctuations
  $\zeta$ (solid lines) as a function of $\zeta/\chizero^2$ for
  various values of the rescaled drift coefficient $\alphahat$, defined in
  \cref{eq:alphahatdef}. The dashed lines are the analytical
  approximation of \cref{eq:Pzetainf}, obtained in the large diffusion
  limit. The tails are asymptotically exponentials. See
  \cref{fig:pzeta} for a zoom within the domain of small
  $\abs{\zeta}/\chizero^2$.}
\label{fig:logpzeta}
\end{center}
\end{figure}

\begin{figure}
\begin{center}
  \includegraphics*[width=\onefigw]{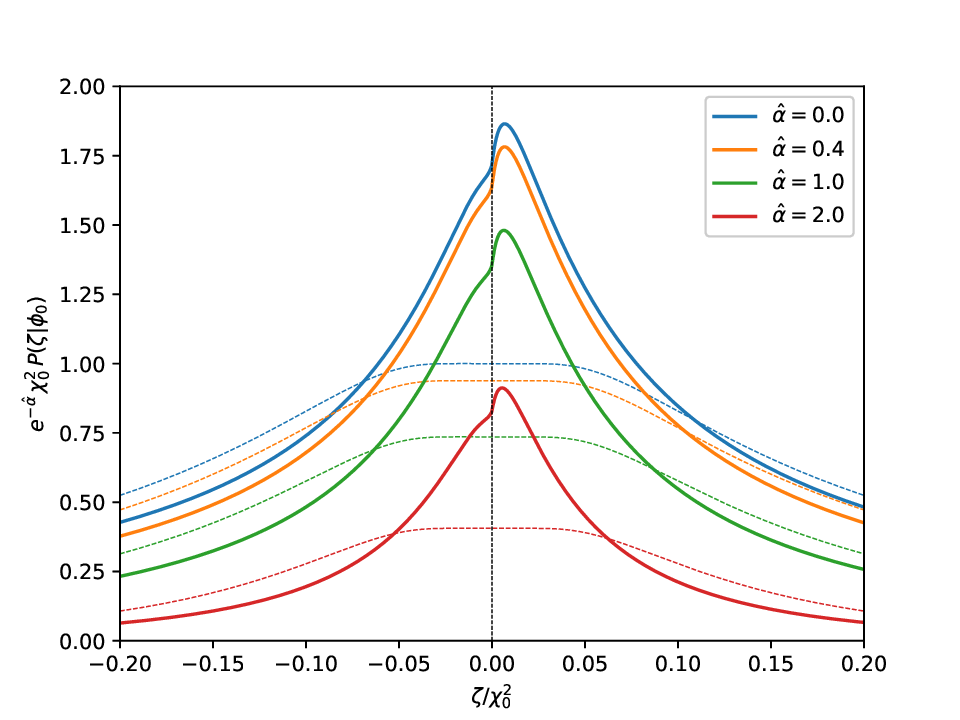}
\caption{Probability distribution of the curvature fluctuations
  $\zeta$ (solid lines) as a function of $\zeta/\chizero^2$ for
  various values of the rescaled drift $\alphahat$. Up to a change
  of amplitude, the shape of the distribution at small curvature is
  similar to the one of the flat semi-infinite potential
  ($\alphahat=0$). It is skewed and the mode is at a positive small
  value which depends on $\alphahat$ (see \cref{fig:mode}). Despite the
  asymmetry, one has $\ev{\zeta}=0$. The dashed lines are the
  analytical approximation of \cref{eq:Pzetainf} providing a good fit
  of the tails (see \cref{fig:logpzeta}).}
\label{fig:pzeta}
\end{center}
\end{figure}

Before turning to the numerical integration of \cref{eq:Pzetanice},
let us derive an analytical approximation for the diffusion
regime. This is the limit $\chizerohat \ll 1$, which will be enforced
as soon as $\abs{\zeta}/\chizero^2 \gg 1$, i.e., in the tails of the
probability distribution for $\zeta$. Plugging
\cref{eq:Pzetatauhalf} into \cref{eq:Pzetanice}, defining the new
integration variable $z=\chizerohat^2/\chizero^2=1/\rv{\Nzero}$, one
gets
\begin{equation}
\Pof{\zeta|\phizero} \underset{\abs{\zeta} \gg \chizero^2}{=} \dfrac{\chizero e^{\alphahat}}{\sqrt{2\pi}}
\, \calI_1\negthickspace\left(x = \dfrac{1}{2\abs{\zeta}},\beta = \dfrac{\alphahat^2}{2
  \chizero^2},\gamma = \dfrac{\chizero^2}{2}\right),
\label{eq:PzetaIone}
\end{equation}
where
\begin{equation}
\calI_n(x,\beta,\gamma) \equiv \int_0^x z^{n-\frac{1}{2}}\, e^{-\frac{\beta}{z} - \gamma z} \dd{z}.
\label{eq:calIndef}
\end{equation}
The family of definite integrals $\calI_n(x,\beta,\gamma)$ can be
analytically calculated from another family of known integrals (see
appendix B of Ref.~\cite{Blachier:2025tcq})
\begin{equation}
I_n(\beta,\gamma) \equiv \int_0^1 z^{n-\frac{1}{2}}\, e^{-\frac{\beta}{z} - \gamma z} \dd{z}.
\label{eq:Indef}
\end{equation}
A change of variable gives
\begin{equation}
\calI_n(x,\beta,\gamma) = x^{n+\frac{1}{2}}
I_n\negthickspace\left(\dfrac{\beta}{x},\gamma x\right),
\end{equation}
from which one gets~\cite{gradshteyn2007}
\begin{equation}
  \begin{aligned}
\calI_1(x,\beta,\gamma)  = -\dfrac{\sqrt{x} e^{-\frac{\beta}{x} -
    \gamma x}}{\gamma} & + \dfrac{\sqrt{\pi} \left(1+2\sqrt{\beta
    \gamma}\right)}{4 \gamma\sqrt{\gamma}} e^{-2\sqrt{\beta\gamma}}
\erfc\left({\sqrt{\dfrac{\beta}{x}}-\sqrt{\gamma x}}\right) -\\ & 
\dfrac{\sqrt{\pi} \left(1-2\sqrt{\beta\gamma}
  \right)}{4\gamma\sqrt{\gamma}} e^{2\sqrt{\beta\gamma}}
\erfc\left(\sqrt{\dfrac{\beta}{x}} + \sqrt{\gamma x}\right),
\end{aligned}
\end{equation}
where $\erfc(\phantom{\cdot})$ is the complementary error function. Plugging this
expression into \cref{eq:PzetaIone}, one finally obtains
\begin{equation}
  \begin{aligned}
  \chizero^2 e^{-\alphahat} \Pof{\zeta|\phizero} &
  \underset{\abs{\zeta}\gg \chizero^2}{=}
  \dfrac{\alphahat+1}{2} e^{-\alphahat} \erfc\left(\alphahat
  \dfrac{\sqrt{\abs{\zeta}}}{\chizero} -
  \dfrac{\chizero}{2\sqrt{\abs{\zeta}}}\right) +
  \dfrac{\alphahat-1}{2} e^{\alphahat}
  \erfc\left(\alphahat\dfrac{\sqrt{\abs{\zeta}}}{\chizero} +
  \dfrac{\chizero}{2\sqrt{\abs{\zeta}}}\right) \\ & -
  \dfrac{1}{\sqrt{\pi}} \dfrac{\chizero}{\sqrt{\abs{\zeta}}}
  e^{-\frac{\chizero^2}{4 \abs{\zeta}} - \alphahat^2 \frac{\abs{\zeta}}{\chizero^2}}.
  \end{aligned}
\label{eq:Pzetainf}
\end{equation}
A consistency check can be made by considering the limit of vanishing drift,
$\alphahat = 0$, for which we recover, as it should, the large lifetime limit for
the flat semi-infinite potential~\cite{Blachier:2025tcq}
\begin{equation}
\chizero^2 \Pof{\zeta|\phizero} \underset{\abs{\zeta}\gg
  \chizero^2}{=}
  \erf\left(\dfrac{\chizero}{2\sqrt{\abs{\zeta}}}\right) -
  \dfrac{\chizero}{\sqrt{\abs{\zeta}}}
  \dfrac{e^{-\frac{\chizero^2}{4\abs{\zeta}}}}{\sqrt{\pi}}\,.
\end{equation}

\begin{figure}
\begin{center}
  \includegraphics*[width=\onefigw]{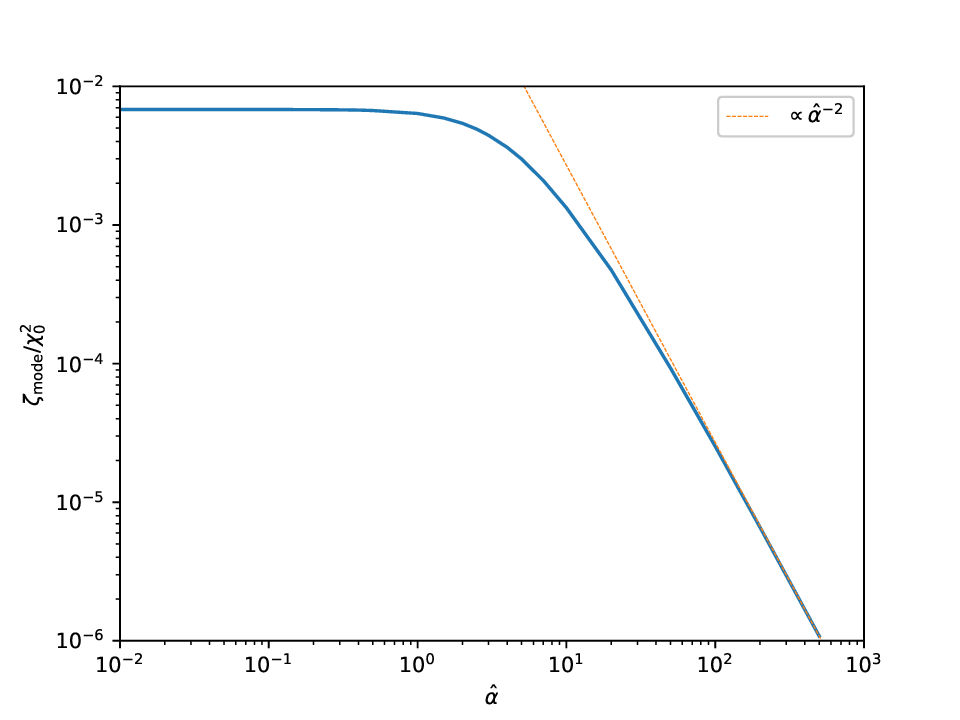}
\caption{Numerical evaluation of the curvature at maximum probability,
  plotted as a function of the rescaled drift coefficient
  $\alphahat$. At small $\alphahat$, $\zeta_\umode/\chizero^2\simeq
  6.9\times 10^{-3}$, which is the value obtained for the flat
  semi-infinite potential without friction. At large values,
  $\zeta_\umode\to 0$ with a power law behaviour in
  $\alphahat^{-2}$.}
\label{fig:mode}
\end{center}
\end{figure}

In \cref{fig:logpzeta}, we have plotted $\Pof{\zeta|\phizero}$ as a
function of $\zeta/\chizero^2$, for various values of $\alphahat$,
obtained from a numerical integration of \cref{eq:Pzetanice} (solid
lines). The large diffusion approximation, \cref{eq:Pzetainf}, is
represented as dashed lines and both distributions are
undistinguishable in the tails. The probability distribution for
$\zeta$ in the region of small $\zeta/\chizero^2$ is represented in
\cref{fig:pzeta}, for various values of $\alphahat$ (solid lines). As
expected, up to a change in amplitude, the shape is mostly the same as
for the flat semi-infinite potential ($\alphahat=0$). In particular,
the distribution is skewed and the mode $\zeta_\umode/\chizero^2$ is
at a positive small value which depends on $\alphahat$. The dependence
of $\zeta_\umode/\chizero^2$ with respect to $\alphahat$ has been
computed and is represented in \cref{fig:mode}.  It goes to zero in
the large drift limit as $\alphahat^{-2}$. Notice that, even though
the large diffusion approximation does not capture these features, the
width of its plateau matches the typical spread of the exact
distribution (see dashed curves in \cref{fig:pzeta}).

In order to better understand the behaviour of $\Pof{\zeta|\phizero}$
in the tails, we can further expand \cref{eq:Pzetainf} in the limit of
$\alphahat \sqrt{\abs{\zeta}}/\chizero \gg 1$. After some tedious algebra, one
obtains\footnote{Expanding the ${\erfc(z)}$ functions of
\cref{eq:Pzetainf} at large $z$, while paying attention to keep the
dependence in $\chizero/\sqrt{\abs{\zeta}}$ in the arguments, all
terms up to order three cancel. The expansion has to be pushed up to
the fifth order to get a non-vanishing leading order term (a
well-known computer algebra software fails in doing this expansion
correctly).}
\begin{equation}
\chizero^2 e^{-\alphahat} \Pof{\zeta|\phizero} \simeq
\dfrac{3}{16\sqrt{\pi}\, \alphahat^2}
\left(\dfrac{\chizero}{\sqrt{\abs{\zeta}}}\right)^5
e^{-\alphahat^2  \frac{\abs{\zeta}}{\chizero^2}}.
\label{eq:Pzetaasymp}
\end{equation}
As such, the asymptotic behaviour of the curvature distribution is a
decaying exponential multiplied by the power-law
$\abs{\zeta}^{-5/2}$. Contrary to the flat semi-infinite case, the
presence of the exponential factor ensures that all moments of
$\Pof{\zeta|\phizero}$ are finite.

\section{Discussion}
\label{sec:discuss}

\subsection{Comparison with the forward formalism}

\begin{figure}
\begin{center}
  \includegraphics*[width=\onefigw]{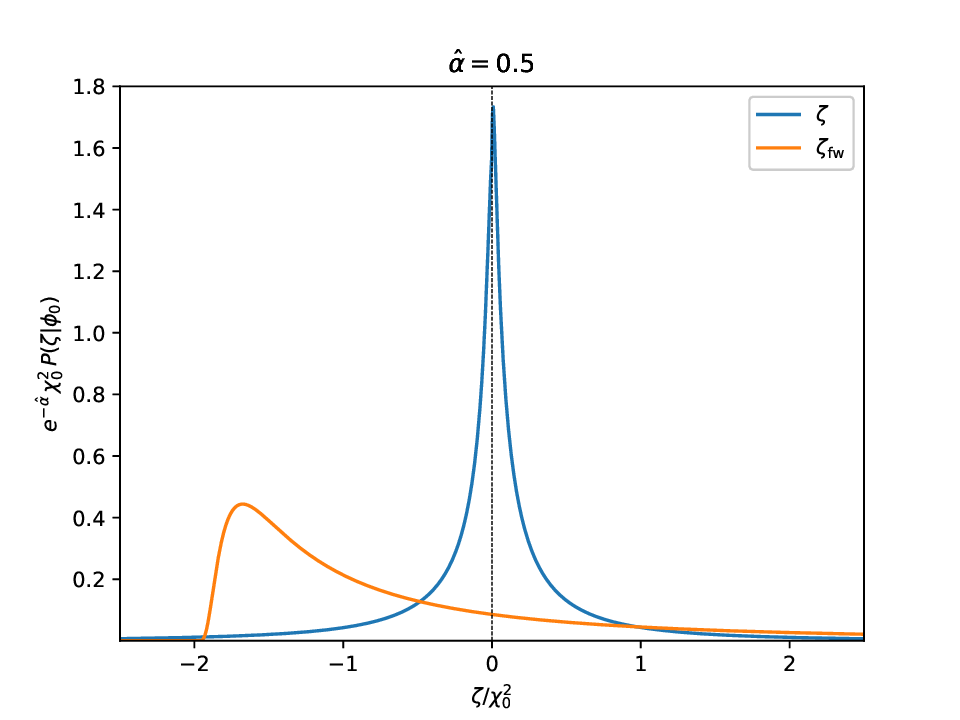}
\caption{Comparison between the curvature probability distribution
  obtained from the time-reversed formalism (blue curve) and the one
  derived from the forward formalism (orange curve), for
  $\alphahat=0.5$. The positive tails are power-law times exponential
  decay but with different exponents (see text). Notice that
  $\Pof{\zetafw|\phizero}$ is exactly vanishing at
  $\zetafw/\chizero^2=-1/\alphahat$ and undefined below.}
\label{fig:Pzetafw}
\end{center}
\end{figure}

In the forward stochastic $\delta N$-formalism, the curvature
fluctuations are defined in reference to the mean lifetime, i.e.,
$\zetafw = \calN- \ev{\calN}$ where $\calN$ denotes the stochastic
variable having for realisations the lifetimes $\rv{\Nzero}$. As
highlighted in the Introduction, without the friction term $\ev{\calN}
= \infty$ and the forward formalism cannot be applied. However, in the
setup considered in the present paper, the constant drift term
regularises the divergence and enables a comparison of the
probability distribution for $\zeta$ obtained between the forward and
the time-reversed formalism.

Recapping that $\Plt{\rv{\Nzero}|\phizero}$ can also be
interpreted as the first passage time distribution for the forward
process, the average of $\calN$ is
\begin{equation}
  \ev{\calN} = \int_0^{+\infty} \rv{\Nzero} \Plt{\rv{\Nzero}|\phizero}
  \dd{\rv{\Nzero}} = \dfrac{\chizero}{\sqrt{2\pi}} \int_0^{+\infty}
  \dfrac{1}{\sqrt{\rv{\Nzero}}}
  e^{-\frac{1}{2}\left(\frac{\chizero}{\sqrt{\rv{\Nzero}}}- \frac{\alphahat
  \sqrt{\rv{\Nzero}}}{\chizero} \right)^2}\dd{\rv{\Nzero}}.
\label{eq:meancalN}
\end{equation}
The same change of variable as in \cref{sec:marge}, $z =
1/\rv{\Nzero}$, allows us to recast this integral as
\begin{equation}
\ev{\calN} = \dfrac{\chizero e^{\alphahat}}{\sqrt{2\pi}}\,
I_0\negthickspace\left(\frac{\chizero^2}{2},\frac{\alphahat^2}{2\chizero^2}\right) = \dfrac{\chizero^2}{\alphahat}\,,
\label{eq:calNmean}
\end{equation}
where $I_0(\beta,\gamma)$ is defined in \cref{eq:Indef}. From
$\Pof{\zetafw|\phizero} = \Plt{\zetafw + \ev{\calN}|\phizero}$, the
forward stochastic $\delta N$-formalism gives
\begin{equation}
\chizero^2 e^{-\alphahat} \Pof{\zetafw|\phizero} =
\dfrac{\alphahat^{3/2}}{\sqrt{2\pi}
  \left(1+\alphahat\dfrac{\zetafw}{\chizero^2} \right)^{3/2}} \, e^{-\frac{\alphahat}{2}\left(1
  + \frac{1}{1 + \alphahat \frac{\zetafw}{\chizero^2}} \right)}
e^{-\alphahat^2 \frac{\zetafw}{2 \chizero^2}}.
\label{eq:Pzetafw}
\end{equation}
In the positive limit $\zetafw/\chizero^2 \gg 1$, the
probability distribution behaves as
\begin{equation}
\chizero^2 e^{-\alphahat} \Pof{\zetafw|\phizero} \simeq
\dfrac{1}{\sqrt{2\pi}}\left(\dfrac{\zetafw}{\chizero^2}\right)^{-3/2} e^{-\alphahat^2
  \frac{\zetafw}{2\chizero^2}},
\end{equation}
a decaying exponential multiplied by a power-law. The exponential term
is actually identical to the one of the tilted
quantum-well in the so-called ``wide'' limit~\cite{Pattison:2017mbe, Ezquiaga:2019ftu,
  Animali:2024jiz} whereas the power-law factor $\zetafw^{-3/2}$ comes
from the fact that the potential is, here, unbounded\footnote{For
unbounded potentials, the poles of the characteristic function are not
forming a discrete set as they do in the bounded case. As such, one cannot
find pure exponential decaying behaviour for the associated
probability density distribution.}.

However, the power-law index is $-3/2$ instead of $-5/2$ found in the
time reverse approach. Also, the exponential decay rate has a factor
of two difference compared to the one of \cref{eq:Pzetaasymp}. Another
important difference is the strong asymmetry of \cref{eq:Pzetafw} with
respect to the sign of $\zetafw$. Indeed, for $\zetafw<0$,
$\Pof{\zetafw|\phizero}$ behaves like a bump function, it is exactly
vanishing at $\zetafw/\chizero^2=-1/\alphahat$ and undefined
below. Indeed, from the very definition of $\zetafw$, $\calN$ being a
positive quantity, one always has $\zetafw \ge -\ev{\calN}$. The mode of
$\Pof{\zetafw|\phizero}$ is at a negative value, which can be quite
large for $\alphahat$ small:
\begin{equation}
\eval{\dfrac{\zetafw}{\chizero^2}}_{\umode}=
-\dfrac{6}{\alphahat\left(3+2\alphahat + \sqrt{9 + 4\alphahat^2}\right)}\,.
\end{equation}
Both $\Pof{\zeta|\phizero}$ and $\Pof{\zetafw|\phizero}$ are
represented in \cref{fig:Pzetafw}. Therefore, for the case of a flat
unbounded potential with constant friction, the time-reversed and
forward approaches do not predict the same curvature distribution,
even in the tails.

This is not really surprising. The time-reversed formalism enforces a
conditioning by the lifetime first, which defines our background as
being one of its possible quantum realisations. As discussed in the
introduction, this conditioning actually sets what are the
cosmological observers in a quantum system. They are the ones
measuring a total amount of accelerated expansion, i.e. a realisation
of $\calN$ having for value $\rv{\Nzero}$. We use the $\delta
N$-formalism only to derive the curvature fluctuations at given
lifetime, namely $\Pof{\zetahat|\phizero,\rv{\Nzero}}$. Then, in a
last step, the curvature fluctuations are weighted by the probability
to be one of these observers, namely, by the probability of the
lifetimes given in \cref{eq:Pltsto}.

Taking average and marginalising over lifetimes do not commute, and
that is why the time-reversed approach leads to a finite result for
$\alphahat=0$ whereas the forward formalism becomes ill-defined. The
forward approach assumes that all of the curvature fluctuations are
generated from the randomness of the lifetimes and the distribution of
$\Pof{\zetafw|\phizero}$ is actually only mapping
$\Plt{\rv{\Nzero}|\phizero}$, up to a constant shift. From
\cref{eq:PzetahatGivenLT,eq:Pzeta}, one can actually reinterpret the
forward result within the time-reversed formalism as postulating a
precise relation for $\Pof{\zetahat|\phizero,\rv{\Nzero}}$ which would
be
\begin{equation}
\Pof{\zetafwhat|\phizero,\rv{\Nzero}} \equiv \dirac{\zetafwhat - 1
  +\dfrac{\ev{\calN}}{\rv{\Nzero}}}.
\end{equation}
Such a distribution would imply that no quantum fluctuations has to be
present at given lifetime, which is not the case for the flat
potential under scrutiny here, as evident in
\cref{fig:Prvdiff,fig:Prvflux}. We should nevertheless stress that,
even if the non-vanishing drift term allows us to regularise the
forward formalism, the system remains highly non-classical because the
potential is still exactly flat. As discussed in the introduction, and
in a way similar to the bispectrum measured by local observers, it is
very well possible that the differences between the time-reversed and
the forward formalism would become negligible in a classical
background. However, at the time of this writing, we do not have solutions
of time-reversed stochastic inflation in non-flat, or even bounded,
potentials.

\subsection{Infinite drift limit}

Although a proper classical limit cannot be defined within a flat
potential, quantum fluctuations at given lifetime happen to be much
reduced in the fluxing regime (see \cref{sec:revsto}). It is possible
to force a very large fluxing by taking the limit of infinite drift,
i.e., $\alphahat\to+\infty$. Indeed, \cref{eq:Pzetanice} also reads
\begin{equation}
\Pof{\zeta|\phizero} = \dfrac{1}{\chizero^2} \sqrt{\dfrac{2}{\pi}}
\int_0^{\frac{\chizero}{\sqrt{\abs{\zeta}}}}
\Pof{\left.\dfrac{\zeta}{\chizero^2} \chizerohat^2\right|\chizerohat}
\chizerohat^2 e^{-\frac{\left(\chizerohat^2 - \alphahat\right)^2}{2\chizerohat^2}},
\label{eq:Pzetanice2}
\end{equation}
showing that, for $\alphahat \gg 1$, the exponential term is acting as
a window function which is non-vanishing only for
$\chizerohat^2=\alphahat$. One can use either a saddle point, or a
Dirac distribution, approximation for the exponential to obtain
\begin{equation}
\Pof{\zeta|\phizero} \simeq \dfrac{\alphahat}{\chizero^2}
\Pof{\left.\zetahat=\alphahat \dfrac{\zeta}{\chizero^2}\right|\chizerohat = \sqrt{\alphahat}},
\end{equation}
provided $\abs{\zeta} < \alphahat^2 \chizero^2$ (and zero
otherwise). Because $\chizerohat = \sqrt{\alphahat}$, the infinite
drift limit implies that we must also consider the infinite fluxing
limit for the curvature distribution at given lifetime, i.e.,
\cref{eq:Pzetahatfluxing}. We therefore get a distribution with
Gaussian tails
\begin{equation}
\lim_{\alphahat \to +\infty}\Pof{\zeta|\phizero} = \dfrac{1}{2
  \abs{\zeta}} e^{-2 \frac{\alphahat^3}{\chizero^4} \zeta^2}.
\label{eq:Pzetainfdrift}
\end{equation}
In contrast, keeping only the dominant terms, the distribution of
$\zetafw$ in the same limit asymptotes to a one-sided exponential
distribution. From \cref{eq:Pzetafw}, one obtains
\begin{equation}
\lim_{\alphahat \to +\infty}\Pof{\zetafw|\phizero} =
\dfrac{\chizero^3 \, \heaviside{\zetafw}}{\sqrt{2\pi} \zetafw^{3/2}} e^{-\frac{\alphahat^2}{2\chizero^2} \zetafw},
\end{equation}
which still has the same exponential tails as the large
$\zetafw/\chizero^2$ limit, and remains quite different from
\cref{eq:Pzetainfdrift}. The Heaviside distribution arises from the
vanishing of $\ev{\calN}$ when $\alphahat \to +\infty$, as explicit in
\cref{eq:calNmean}. The infinite drift therefore acts as a
classical-like limit but only for the reverse process. Once more, this
is not surprising as the forward formalism is not sensitive to the
fluctuations at given lifetime and, therefore, $\zetafw$ cannot be
affected by the quantum fluctuations occurring, or not, at a given
realisation of $\calN$.

\subsection{Negative drift}

In order to further illustrate the differences between the forward and
time-reversed approaches, we can consider another quite extreme case
where the drift $\alphahat < 0$. In the regime where the friction term
is approximately modelling a slightly tilted potential, as in
\cref{eq:tiltpot}, a negative drift would represent a potential
tilted down at larger field values, allowing for field trajectories to never reach
the end of quantum diffusion. Indeed, from \cref{eq:survival,eq:Pltsto}, 
one can derive the survival probability for
non-vanishing friction. Its asymptotic limit reads
\begin{equation}
\lim_{N\to\infty} S(N) = \heaviside{-\alphahat} \left(1 -
e^{-2|\alphahat|} \right).
\end{equation}
It is non-vanishing for $\alphahat < 0$, and, for large negative
values, only an exponentially suppressed fraction of all the
realisations succeeds in crossing $\phi=\phiqw$.

For the forward formalism, such a
situation is more pathological than the flat semi-infinite potential:
the drift is pushing the field $\phi$ away from $\phiqw$. For
instance, \cref{eq:calNmean} implies that the mean number of (forward)
{\efolds} is negative whereas \cref{eq:Pzetafw} is only defined for
$\zetafw > -\chizero^2/\alphahat$, which can be very large for small
negative $\alphahat$. This would suggest that only very large positive
curvature fluctuations can be generated for an infinitely small and
negative drift.

For time-reversed stochastic inflation, \cref{eq:Pzetanice} shows that
only the term in $e^{\alphahat}$ behaves differently than in the case
where $\alphahat >0$, all other terms depending on $\alphahat^2$. As
such, $\Pof{\zeta|\phizero}$ remains well-defined, has the same
exponential tails as in \cref{eq:Pzetaasymp}, and, only the overall
amplitude is reduced by the $e^{\alphahat}$ factor. However,
\cref{eq:Pzetanice2} shows that, for negative drift, the Gaussian
limit no longer exists. As discussed at length in
Ref.~\cite{Blachier:2025tcq}, all of this is expected. The
conditioning by the lifetimes ensures that we only count realisations
in which inflation has ended. For negative drift, many realisations
are actually inflating forever and never reach the quantum wall. But
the time-reversed formalism only picks up the trajectories reaching
the wall, which explains the finiteness of the curvature distribution,
as well as the presence of the overall exponential damping factor
$e^{\alphahat}$ in \cref{eq:Pzetanice}. This factor may actually be
interpreted as coming from the quantum tunneling through the potential
barrier set by $\alphahat<0$~\cite{Animali:2022otk}. Let us also stress
that, for $\alphahat<0$, $\Pof{\zeta|\phizero}$ is no longer normalised
to unity but to $e^{-2|\alphahat|}<1$, which is precisely
$1-S(\infty)$, the overall probability that stochastic inflation
ends. Finally, the realisations escaping the quantum region must be
driven by quantum kicks, jumping over the stream set by
$\alphahat<0$. It justifies why the Gaussian limit exhibited in
\cref{eq:Pzetainfdrift} cannot exist anymore, since it is precisely
the diffusion that has to dominate in order to produce the kicks
required for inflation to end.

\section{Conclusion}
\label{sec:conc}

In this work, we have solved time-reversed stochastic inflation for the
flat semi-infinite potential in presence of a constant drift term. The
stochastic $\delta N$-formalism, applied to the reverse process, has
allowed us to derive the probability distribution
$\Pof{\zeta|\phizero}$ of the curvature fluctuations in
\cref{eq:Pzetanice}, which is plotted in
\cref{fig:logpzeta,fig:pzeta}. The drift term, encoded in the
parameter $\alphahat$, appears to regularise the tails of the
distribution $\Pof{\zeta|\phizero}$. It exhibits an exponential decay
which contrasts to the Levy-like power-law tails studied in
Ref.~\cite{Blachier:2025tcq} for the driftless case
$\alphahat=0$. Considering the limit of infinite drift, we have found
in \cref{eq:Pzetainfdrift} that $\Pof{\zeta|\phizero}$ develops
Gaussian tails, which are typical of semi-classical processes. The
large drift limit indeed corresponds to the large fluxing limit of the
reverse process for which quantum fluctuations at given lifetime are
much reduced.

Another interest of considering a constant drift term, in a flat
semi-infinite potential, is that it regularises the forward formalism
which is otherwise ill-defined. The curvature distribution is there
identified with the fluctuations of the lifetime and
$\Pof{\zetafw|\phizero}$ has been derived in \cref{eq:Pzetafw}. Up to
some similarity with the time-reversed result, as having exponential
tails, it is quantitatively different, as apparent in
\cref{fig:Pzetafw}. It is strongly asymmetric, behaving like a bump
function for negative curvatures, and does not reach a Gaussian-like
distribution in the infinite drift limit. As discussed in the
introduction, both approaches differ by counting the stochastic number
of {\efolds} either forward, from the initial state $\phizero$ to the
absorbing wall at $\phiqw$, or reverse, from the final state $\phiqw$
towards the initial state $\phizero$. In the semi-classical limit, and
slow-roll, it is known that both methods can lead to results differing
by first order slow-roll corrections, which are therefore always
small. Here slow-roll does not exist, the potential being exactly
flat, quantum kicks are the driving dynamics of the system and they
are only tampered by the added friction term. It is therefore not
surprising that we find quite different results between the forward
and reverse picture. It is possible that these differences become
sub-dominant in more classical setups. Obtaining solutions for
time-reversed stochastic inflation in more classical systems than the
one presented in this work would be a way to confirm this hint, albeit
technically challenging.

Let us also stress that the exponential positive tail of
$\Pof{\zeta|\phizero}$ and $\Pof{\zetafw|\phizero}$ happens to differ
by a factor of two in the exponent. Such a factor may be quite
important would we want to predict abundances of PBH in another
context. In the time-reversed picture, one can trace this factor as
coming from the fact that curvature fluctuations, at given lifetime
$\rv{\Nzero}$, are in reference to $\ev{\rv{N}}$. As such, they have a
tendency to be centered on $\rv{\Nzero}/2$, as can be seen in
\cref{fig:Prvdiff}. On the contrary, in the forward picture,
fluctuations at given lifetime are not accounted for, the lifetime
alone is the main stochastic quantity and the curvature fluctuations
are in reference to $\ev{\calN}=\ev{\rv{\Nzero}}$. This difference in
the reference point is at the origin of this factor of two difference
in the exponential decay rates. It it also at the origin of the strong
asymmetry of $\Pof{\zetafw|\phizero}$ with respect to the sign of
$\zetafw$. As a result, the choice of $\Nflat$ in the definition of
the stochastic $\deltaN$-formalism is not innocuous when quantum
diffusion dominates.

Finally, we have shown how a negative drift leads to finite results in
the time-reversed approach while being pathological in the forward
formalism. Such a negative drift is actually a simple case of eternal
inflation in which many field trajectories never reach the end of
quantum diffusion. Our findings illustrate how the conditioning by the
lifetimes, hardcoded in the time-reversed stochastic inflation
formalism, enforces the regularity of the extracted probability
distributions. In more intuitive terms, why should we be concerned
with trajectories in which inflation never ends? Assessing the
statistics of cosmological observables should be performed in the
subset of all possibilities leading to our universe, i.e., the
Friedmann-Lema\^{i}tre decelerating models. It seems to us that, in
the absence of a well-defined classical background, considering all of
its possible quantum realisations is the sensible alternative.

\section*{Acknowledgements}
We warmly thank C.~Animali and P.~Auclair for comments and suggestions
on the first version of this manuscript as well as E.~Tomberg for
enlightening discussions. This work is supported by the ESA Belgian
Federal PRODEX Grants $\mathrm{N^{\circ}} 4000143201$ and
$\mathrm{N^{\circ}} 4000144768$. B.~B. is publishing in the quality of
ASPIRANT Research Fellow of the FNRS.

\appendix

\bibliographystyle{JHEP}

\bibliography{references}

\end{document}